\documentstyle[12pt,aasms4]{article}
\begin{document}

\begin{center}
\title{$Spitzer$  IRS Spectroscopy of Intermediate Polars: Constraints on
Mid-Infrared Cyclotron Emission}

\author{Thomas E. Harrison$^{\rm 1}$, Ryan K. Campbell$^{\rm 1}$}

\affil{Department of Astronomy, New Mexico State University, Box 30001, MSC 4500, Las Cruces, NM 88003-8001}

\authoremail{tharriso@nmsu.edu, cryan@nmsu.edu}

\author{Steve B. Howell}

\affil{WIYN Observatory and National Optical Astronomy Observatories, 950 North
Cherry Avenue, Tucson, AZ 85726}

\authoremail{howell@noao.edu}

\author{France A. Cordova}

\affil{Institute of Geophysics and Planetary Physics, Department of Physics,
University of California, Riverside, CA 92521}

\authoremail{france.cordova@ucr.edu}

and

\author{Axel D. Schwope}

\affil{Astrophysikalisches Institut Potsdam, An der Sternwarte 16, D-14482
Potsdam, Germany}

\authoremail{aschwope@aip.de}

\end{center}

$^{\rm 1}$Visiting Astronomer at the Infrared Telescope Facility, which is
operated by the University of Hawaii under contract from the National
Aeronautics and Space Administration.\\

\noindent
{\it Key words:} infrared: stars ---  stars:
individual (AE Aquarii, FO Aquarii, V603 Aquilae, TV Columbae, TX Columbae, DQ 
Herculis, EX Hydrae, PQ Geminorum, GK Persei, AO Piscis, V1223 Sagitarii)

\abstract{
We present $Spitzer$  Infrared Spectrograph (IRS) observations of eleven
intermediate polars (IPs). Spectra covering the wavelength range from 5.2
to 14 $\mu$m are presented for all eleven objects, and longer wavelength spectra
are presented for three objects (AE Aqr, EX Hya, and V1223 Sgr).
We also present new, moderate resolution (R $\sim$ 2000) near-infrared spectra 
for five of the program objects.  We find that, in general, the mid-infrared 
spectra are consistent with simple power laws that extend from the optical into 
the mid-infrared. There is no evidence for discrete cyclotron emission features
in the near- or mid-infrared spectra for any of the IPs investigated.
Binning the spectra to create photometry, we find no evidence for infrared
excesses at $\lambda$ $\leq$ 12 $\mu$m. However, AE Aqr, and possibly EX Hya and V1223 
Sgr, show longer wavelength excesses. The lack of discrete cyclotron features
suggests that if there is any such emission in the bandpass of the IRS, it must
be from optically thick cyclotron emission. We have used a cyclotron modeling 
code to put limits on the amount of such emission for magnetic field 
strengths of 1 $\leq$ B $\leq$ 7 MG. For those objects with high quality 
spectra, we find that any optically thick cyclotron emission must be less than 
50\% of the local continum.
If cyclotron emission is occurring in the 5.2 to 14.0 $\mu$m bandpass it
constitutes less than 1\% of the bolometric luminosity of $any$ of the
IPs. We were able to model the long wavelength excess of V1223 Sgr and EX Hya 
with cyclotron emission from a 1 MG field, but the S/N of those data are very 
poor, and the reality of those excesses is not established. We attempted to 
model the long wavelength excess of AE Aqr with cyclotron emission, but 
none of our models fit nearly as well as a simple, cool (T$_{\rm BB}$ = 140 K)
blackbody.  Given the apparent variability of this excess, however, synchrotron 
radiation remains a better explanation. We discuss our results in the context of 
the standard model for IPs.
}

\section{Introduction}

Intermediate polars (IPs) are a subclass of cataclysmic variable that are 
believed to harbor magnetic white dwarf primaries that are accreting matter 
from low mass, late-type secondary stars. They are identified by a combination
of multi-periodic photometric behavior, and hard X-ray spectra (see Warner 
1995). They differ 
from the ``polars'' in that the magnetic field strength of the primary is 
believed to be weaker, based on the lack of detectable polarization or 
discrete cyclotron harmonics in their optical/near-infrared spectra. Also, in 
contrast to polars, the white dwarf 
primaries in IPs are rotating asynchronously with periods that are substantially
shorter than the orbital period of the binary. Instead of capturing the 
accretion stream close to the secondary star, and funneling it onto the 
magnetic poles, a truncated accretion disk appears to be present in most IPs, 
the inner edge of which is then captured by the magnetic field of the white 
dwarf, though ``stream-fed'' systems are believed to exist (e.g., Norton et al.
2004a).

The rapid rotation of the white dwarf creates a magnetosphere, inside of which
the accreted matter flows along the field lines and accretes onto the
primary after passing through a shock. The shock is the source of the X-ray
emission, and the spin rates of the white dwarfs in many IPs have been
discovered using X-ray observations. Due to the possiblity 
of either gaining or losing angular momentum through the interaction of the 
magnetic field with the accreting matter, the white dwarfs in IPs could be 
either spun-up (e.g., AO Psc, van Amerongen et al. 1985) or spun-down
(e.g., V1223 Sgr, van Amerongen et al. 1987). But Norton et al. (2004b) show that 
there is a large range of parameter space that allows for rotational 
equilibrium.

In the synchronously rotating polars, the accretion process is simpler, and as the material approaches
the white dwarf primary, a hot shock forms, emitting X-rays. Depending on
the accretion rate, cyclotron emission can be the dominant cooling process.
During periods of relatively low mass accretion rates, strong, discrete
cyclotron harmonics are observed in optical and/or infrared spectra of polars 
(Fig. 1), that can lead to robust estimates of the magnetic field strength, and
allow for estimations of the conditions in the accretion column.
The presence of 
X-ray emission and, in a few cases, circularly polarized light, suggests that 
magnetically controlled accretion $is$ occurring in IPs. However, most IPs
do not show any sign of polarization, and no IP has ever displayed discrete
cyclotron emission humps. Thus, a standard model has been constructed which 
assumes that the field strength of the magnetic white dwarf primary must be 
lower than B $\leq$ 8 MG, and that the accretion rates in IPs are higher than
those in polars. Wickramasinghe et al. (1991) show that this model is 
consistent with the observational data for IPs. 

Due to higher accretion rates, discrete cyclotron emission is suppressed, and 
any circular polarization from the higher harmonics is highly diluted by the 
flux from the disk. Thus, to search for the cyclotron emission from IPs, 
observations
at mid-infrared wavelengths are necessary. At wavelengths longer than 2 $\mu$m,
the cyclotron emission should be detectable for fields with B $\leq$ 10 MG
even in the presence of strong disk emission. Since mid-infrared polarization
is not yet feasible for such faint sources, detection of excess emission using 
mid-infrared spectroscopy may be our only method for confirming this model and 
estimating field strengths, and accretion parameters in IPs.

We present a mid-infrared spectroscopic survey of eleven IPs obtained using the 
Infrared Spectrograph (IRS) on the Spitzer Space Telescope. We find no evidence 
for cyclotron emission at wavelengths below 12 $\mu$m and, in general, the IRS 
spectra are consistent with a simple power-law form extrapolated from the 
optical and near-infrared into the mid-infrared. We do find a significant excess 
at $\lambda$ $>$ 12 $\mu$m for AE Aqr, but conclude that it is also not due to 
cyclotron emission. In the next section we discuss the $Spitzer$ observations, 
we then present our results, and discuss their implications.

\section{Observations}

The Infrared Spectrograph (IRS) on the Spitzer Space Telescope has been
described by Houck et al. (2004). The IRS provides spectra from 5.2 to
38 $\mu$m at resolutions of R $\sim$ 90 and 600. The data presented below
were all obtained in the low resolution mode. At low resolution, there are
four different observing modes, SL1, SL2, LL1, and LL2, that correspond to
four different spectral regions. SL1 uses a 5.84 line/mm grating (blazed
at 11.5 $\mu$m) in first order, covering the wavelength region of 7.4 $\leq$ 
$\lambda$ $\leq$ 14.0 $\mu$m. SL2 uses the same grating in 2nd order, and 
covers the spectral region 5.2 $\leq$ $\lambda$ $\leq$ 7.7 $\mu$m. A similar
grating (5.62 lines/mm blazed at 31.5 $\mu$m) is used in the long wavelength
mode, with LL1 being first order, covering the wavelength region 19.5 to
38 $\mu$m, and LL2 second order, covering the region 14.0 to 23.0 $\mu$m.
To change from one mode to another requires offsetting the telescope to
a slightly different slit position.

A journal of the observations for our nine program IPs is provided in Table 1, 
where we list the observation date, the start time of the observations (UT), the 
observing modes with exposure times and the number of cycles in each mode, the 
total duration of the observations, the IP orbital period, the spin period of 
the white dwarf, and the binary orbital inclination. In addition to the IRS data 
for our nine IPs, we extracted archived IRS observations of AE Aqr (Houck GTO-2 
``IRS\_DISKS''), and GK Per (Gehrz ``RDG-GTO''), two peculiar members of this 
heterogeneous CV subclass. AE Aqr is significantly brighter than any of the 
other IPs, and excellent spectra were obtained in both short and long wavelength 
modes. In addition to the observations listed in Tablei 1, AE Aqr was observed 
for three cycles in the LL1 mode with 100 sec exposure times. GK Per was only 
observed in the SL1 and SL2 modes.

To provide estimated mid-infrared fluxes for planning our observations, we 
simply used a power law extrapolation of
the optical/IR SED of each target. Suprisingly, these estimations were quite
good, falling within 20\% of the observed fluxes. Our goal was to achieve
a S/N = 10 in the continuum for each object in the SL1 and SL2 modes. This was 
not attained for any of our nine sources, however, with typical S/N ratios near 6.
We had hoped for similar quality spectra in the LL2 mode, but only two of
the nine target IPs (EX Hya and V1223 Sgr) were detected in the LL2 mode, and 
both at very low S/N ($\leq$ 1.5). While this result is somewhat discouraging, 
note that the hope was to detect cyclotron emission above, and beyond that of 
the accretion disk
continuum, thus our results provide useful constraints on any mid-infrared 
cyclotron emission.  Due to the expected faintness of our nine targets, we did 
not attempt observations of any of our nine IPs in the LL1 mode. 

Due to the fact that both of the arrays in the instrument are simultaneously
exposed, data on the background are measured in SL1 (LL1) while observing in 
SL2 (LL2), and vice versa. Thus, while we nodded the program objects along the 
slit for sky subtraction (a ``cycle'') in each observing mode, additional sky 
spectra were automatically obtained from the array which contained no source 
spectra. Thus, in some cases (V603 Aql and V1223 Sgr), up to 15 sky spectra were 
obtained that could be medianed together and subtracted
from an object spectrum. Before using the ``Spitzer IRS Custom
Extractor'' (SPICE) tool to extract our program object spectra, we used
IRAF to median combine all of the sky data, and then subtract these from
the program object spectra. This procedure produced a much cleaner sky 
subtraction
than if we had simply subtracted alternating images. The SPICE\footnote{For the
SPICE manual go to http://ssc.spitzer.caltech.edu/postbcd/doc/spice.pdf} package
extracts and flux calibrates the IRS spectra using both bad pixel masks, and
other calibration data that come included with your observations. All of our
extractions were performed using the standard procedure. The final IRS spectra
for ten of the IPs are shown in Fig. 2. The data for AE Aqr are shown in
Fig. 3. The error bars on the fluxes presented
in these figures were determined using IRAF, since SPICE does not properly
propogate the various uncertainties through the reduction process. As such,
these errors are the means for each spectrum and do not reflect the
true uncertainties of any single data point (the errors on the fluxes generally
increase with wavelength in both spectral ranges).

In addition to the IRS data just noted, we present near-infrared spectra
of the polar EF Eri, and five IPs obtained using SPEX (Rayner et al. 2003) on 
the IRTF.  EF Eri was observed in low resolution mode (R $\sim$ 250) on 16 
August 2004.  The IPs V603 Aql, AE Aqr, FO Aqr, DQ Her and GK Per were
observed on 15 August 2004. The 
IPs were observed using the cross-dispersed mode, covering the 
wavelength interval 0.80 $\leq$ $\lambda$ $\leq$ 2.42 $\mu$m, with a
resolution of R $\sim$ 2000. Nearby A0V stars were used for telluric 
correction, and these data were telluric corrected, and flux and wavelength 
calibrated using the SPEXTOOL package (Vacca et al. 2003). The spectrum of EF
Eri is presented in Fig. 1, and those for the IPs are presented in Figs. 4 
through 8.

\section{Results}

As shown in Fig. 1, the cyclotron harmonics in polars are very broad features
that can be easily seen in low resolution spectra. However, no such features
are see in the IRS spectra of $any$ of the IPs. While there is structure 
in the spectra of all of the IPs, we do not believe these
are due to cyclotron features. The best case for ``optically thin'' (see
below) cyclotron emission
appears to be V603 Aql, with a strong emission feature at 8 $\mu$m, but several 
of the other IPs have ``features'' or discontinuties in the wavelength region
between 7 and 9 $\mu$m. There appears to
be some type of flaw in the data, or reduction process, that has difficulty
in this wavelength region for low S/N data (the data for V603 Aql are quite
poor).  While we cannot completely dismiss the possibility that these
features are real, it will take additional observations to confirm their 
existence.  In addition to the 8 $\mu$m region, there is an artifact 
that produces excess emission at wavelengths near 14 $\mu$m in the SL1 mode. 
As described in the IRS Handbook, this ``teardrop'' is apparently due to a 
scattered light problem. The most recent versions (we used version S13) of 
the pipeline processing greatly reduce the effect, but do not appear to 
completely eliminate it from the spectra of very faint sources. Thus, any flux 
excesses beyond 12 $\mu$m in the SL1 mode should not be trusted in the data 
presented here. As shown in Fig. 3, AE Aqr has a significant long wavelength
excess, and we treat it separately, below.

Given the rather low S/N of the IRS data, we decided to create photometry from 
these spectra to better examine the spectral energy distributions (SEDs) of these 
objects for possible mid-infrared excesses. For this process, we created mean 
flux densities of the sources in one micron increments. Because the spectra do 
not exactly start and stop within perfectly defined wavelength intervals, the 
fluxes at the endpoints of each mode are not fully one micron intervals. For those
intervals, we calculated the mean wavelength of the data that was averaged.
We have compiled photometry of all of the IPs from published sources, and
construct SEDs from the $U$-band through 14 $\mu$m for eight sources, and
through 21 $\mu$m for EX Hya and V1223 Sgr. We present these data in Fig. 9. 
Note that the LL2 data at the longest wavelengths 
for both EX Hya and V1223 Sgr generally have a S/N $\leq$ 1, and thus are 
somewhat unreliable, even when averaged.

Given the variable nature of the optical and near-infrared emission from these 
objects, it is remarkable how well non-simultaneous data can be fitted with a 
simple power-law spectrum. The optical/infrared SEDs for V603 Aql, FO Aqr, 
PQ Gem, 
AO Psc, TV Col and V1223 Sgr are easily modeled with power law spectra (f$_{\nu}$
$\propto$ $\nu^{\alpha}$) with indices of 0.8 $\leq$ $\alpha$ $\leq$ 1.3.
Note also that there is no evidence for secondary stars in our IRTF data for
V603 Aql, FO Aqr or DQ Her. The SEDs of TX Col and DQ Her are not as easily fit
with power laws, but this could be due to changes in state between the epochs of
the optical/near-infrared data and the $Spitzer$ observations. The IRS data for
both objects are consistent with power law spectra. The SED for EX Hya suggests
that the secondary star is a significant contributor to the near-infrared
fluxes, while the SED for GK Per is dominated by its subgiant secondary star in 
the optical and near-infrared.

It is clear that the dominant source of luminosity in the shorter period 
IPs is due to their accretion disks.  We find that there is no evidence for 
mid-infrared excesses in these ten IPs. This is in stark contrast to $Spitzer$ 
mid-infrared (IRAC) photometry of several polars (Howell et al. 2006, Brinkworth
et al. 2006). In nearly
every polar observed, a significant mid-infrared excess was detected. While 
the origin for those excesses is unclear, it is possible that a significant 
fraction of the mid-infrared fluxes in polars can be attributed 
to cyclotron emission from the lowest harmonics ($n$ $\leq$ 3.0).

\subsection{Limits on Cyclotron Emission}

To attempt to place limits on the cyclotron emission in the program IPs, we have
generated a large number of models using the one-dimensional cyclotron code 
first developed by Schwope et al. (1990). This code assumes a constant temperature
and plasma ``size parameter'' ($\Lambda$), which is related to the optical
depth of the accretion region. We constructed models for field strengths of 0.5, 
1, 2, 3, 5 and 7 
MG, at two plasma temperatures (T$_{\rm pl}$ = 10 and 20 keV), and for two 
viewing angles ($\Theta$ = 57 and 85$^{\circ}$). We ran models that covered a 
large range in the size parameter: $-$5.0 $\leq$ log($\Lambda$) $\leq$ 7.0. For 
the discussion below, we refer to models with log($\Lambda$) $\leq$ 1.0 to
be optically thin, models with 1.0 $<$ log($\Lambda$) $<$ 5 to be of
``intermediate'' optical depth, and models with log($\Lambda$) $\geq$ 5.0 to
be optically thick.  A run of these models over 
a large range in optical depth for a 3 MG field is shown in Fig. 10a. Note
that as the optical depth increases, the discrete cyclotron harmonics merge
together to form a continuum. To derive constraints on the possible contribution
of the cyclotron emission we combined these models with simple
power law spectra that best fit the optical to mid-IR data for
each object. Figs. 10b,c show an example of the fitting procedure for EX Hya.
Obviously, it is difficult to place tight limits
on the contribution of the cyclotron flux given the quality of the IRS spectra,
and the possibility that the cyclotron emission is optically thick. As shown in
Fig 10b,c, it is easy to rule out cyclotron emission for plasmas
with  log($\Lambda$) $\leq$ 1.0, for field strengths between 3 and 7 MG in 
{\it all} of our sources to very stringent limits: $\leq$ 20\% of the continuum at the 
$n$ = 3 (B $\geq$ 4 MG) or the $n$ = 4 (B $<$ 4 MG) harmonic (selecting the
harmonic that fell closest to the center of the SL1 + SL2 bandpasses). At 
field strengths outside this range, the most prominent cyclotron humps fall
outside the bandpasses of the data presented here.  However, field strengths
larger than 7 MG have harmonics that are located in the near-infrared. If
optically thin cyclotron emission was present, we would have clearly detected it
in our SPEX data for the five IPs surveyed. Thus, if the cyclotron emission from 
IPs is optically thin, it must arise from field strengths of B $<$ 3 MG. 

As the optical depth of the cyclotron region increases, its affect on the 
continuum becomes increasingly difficult to separate from a power law spectrum
(see Fig. 10a).  Thus, even 
for the data with the highest S/N (e.g., EX Hya), we cannot rule out 
contributions of up to 50\% for optically thick cyclotron emission. For those
objects with lower S/N data, we cannot rule out the possibility that their
mid-infrared spectra are dominated by optically thick cyclotron emission.
Even with such limits, however, it is important to realize that this is a
significant constraint. For example, in polars, the optical/IR cyclotron 
emission can be a considerable fraction of their bolometric luminosities. 
While IPs have a 
strong accretion disk component that polars do not, the contrast between the 
cyclotron emission and the local accretion disk flux increases with increasing 
wavelength, due to fact that the peak of the cyclotron emission for low strength 
fields falls in the IRS bandpasses, while the accretion disk spectrum is
declining. 
{\it We conclude that any contribution from optically thick cyclotron emission in the wavelength 
interval 5.2 $\leq$ $\lambda$ $\leq$ 14 $\mu$m is $<$ 0.5 \% of the bolometric 
luminosities of the program IPs.}

V1223 Sgr, and to a lesser extent, EX Hya, both show excesses in the LL2 
bandpass (Fig. 9b). While these spectra are quite poor, most of the photometric 
fluxes in this spectral region have S/N 
$>$ 3.  As discussed for AE Aqr, below, we can model excesses in this wavelength 
range with cyclotron emission from fields with 1 $\leq$ B $\leq$ 2 MG. The final 
model for AE Aqr, derived below, fits the data for both EX Hya and V1223 Sgr 
reasonably well. Given the similarity between the long wavelength excess in 
V1223 Sgr and AE Aqr, it would be useful to obtain higher quality LL2 and new 
LL1 spectra of both V1223 Sgr and EX Hya to conclusively determine the reality of 
their apparent excesses.

\subsection{The Mid-Infrared Excess of AE Aqr}

In contrast to the results above, is the spectrum of AE Aqr. AE Aqr is
a peculiar object. The white dwarf in AE Aqr spins faster than any other
(33 s), but this rotation rate is also rapidly slowing, at a rate of 
6 $\times$ 10$^{\rm -14}$ s s$^{\rm -1}$ (De Jager et al. 1994), the most 
rapid spin-down rate known. This rapid spin down cannot be explained by normal 
accretion torques, and thus a ``magnetic propeller'' model (Wynn et al.  1997) 
has been developed where the 
spinning magnetic field of the white dwarf ejects material from the system. 
It has been suggested by Schenker et al. (2002) that AE Aqr is
a brand new CV, having just emerged from the common envelope phase. UV
spectra of AE Aqr reveal an extreme N V/C IV ratio, suggesting a deficit of
carbon, and an enhancement of nitrogen. This indicates that material which
has been processed through the CNO cycle is making its way to the surface of the
white dwarf primary. The near-infrared spectrum of AE Aqr is consistent
with these results, revealing a late-type (K4) secondary star, {\it but
one that has no evidence for $^{\rm 12}$CO absorption features!} This implies 
a secondary star that is highly deficient in carbon, a general result
found for many CVs (see Harrison et al. 2004a, 2005a), but AE Aqr and GK Per
are the most extreme cases we have found in our infrared spectroscopic surveys.

The IRS spectrum of AE Aqr is shown in Figure 3. Shortward of 12 $\mu$m, the
spectrum is consistent with a power law, but beyond this wavelength there
is a considerable flattening of the spectrum. The SED
of AE Aqr, including the IRS data, is plotted in Fig. 11. 
In addition to the IRS data, 
we plot point source fluxes from IRAS at 25 and 60 $\mu$m, and ground-based
observations at $M$', 11.7 and at 17.6 $\mu$m from Dubus et al. (2004). Dubus 
et al. reported that these fluxes were highly variable. If we ignore the IRS 
data, the 17 through 60 $\mu$m fluxes of AE Aqr 
can be fitted with a cool blackbody (T$_{\rm BB}$ = 125 K). The IRS data
(see Fig. 3), can also be fit with a blackbody, but one with a slightly
higher temperature: T$_{\rm BB}$ = 140 K. Given the likelihood of
matter being ejected from the binary system, a circumstellar dust shell around
AE Aqr is not unexpected. Assuming black grains, this dust intercepts
about 1\% of the bolometric flux from AE Aqr.

Given the evidence for short timescale variability, a circumstellar 
dust shell is probably not the best explanation for the mid-infrared excess
of AE Aqr. Dubus et al. (2004) discuss the possibility 
of explaining their observations with synchrotron radiation, a mechanism 
consistent with both the mid-infrared and radio properties of this source. 
In addition, AE Aqr has been detected at TeV energies (Meintjes et al. 1992)!
Given that there is no apparent turnover in the slope of the spectrum shown
in Fig. 3, suggests that the transition from optically thin to thick 
synchotron emission occurs at wavelengths longer than 38 $\mu$m (see Dubus
et al. 2004).
Here, however, we explore the possiblity that the observed mid-infrared excess 
is due to cyclotron emission, a process that can also explain the rapid 
variability of this source. 
It is clear that well-defined, discrete cyclotron harmonics are not present
in the IRS data for AE Aqr, suggesting that any cyclotron emission must
be optically thick. Our models indicate that the magnetic field strength cannot 
be much higher than 1 MG, as the excess would occur at shorter wavelengths. As 
the optical depth of the cyclotron emission region is
increased, the maximum emission shifts away from the lower harmonics (and
their longer wavelengths), forming a psuedo continuum in the higher harmonics,
at shorter wavelengths. For completely optically thick emission at a field 
strength of 
1 MG, the peak in the cyclotron flux occurs near 10 $\mu$m. To provide
the observed infrared excess requires emission from a region of intermediate
optical depth, and such would have discrete harmonics in the observed bandpass. 
While we have not explored the entire parameter space, we find that
the best fitting model has B = 1 MG, and is of intermediate optical depth (Fig.
11). None of our models could create a featureless excess that is as broad
as that which is observed. Clearly, the blackbody fit does a better job,
and we therefore believe that this excess is probably not due to cyclotron emission.

\section{Discussion}

While there is no {\it a priori} expectation that the magnetically controlled 
accretion in IPs is similar to that in polars, their X-ray properties, and
the detection of circular polarization in some IPs argues that they share some 
commonality. Thus, the complete lack of detectable cyclotron emission in these 
IPs is 
puzzling. There are a number of possible explanations, 1) cyclotron emission is 
only emitted during a small fraction of a spin/orbital period, or does not
occur along our line of sight, 2) the properties of the IPs examined here are 
such that cyclotron emission is unimportant, or 3) the magnetic field 
strengths of IPs are higher or lower than currently believed.
We examine each of these scenarios, below.

Cyclotron emission from most polars {\it is} only visible during part of an 
orbit. For some polars, there are times during an orbital period where the 
cyclotron emission disappears, or at least becomes insignificant,
compared to other sources in the system. In other cases, such as EF Eri (see 
Harrison et al.  2004b), cyclotron emission apparently dominates throughout
an orbital period. It could be that the field geometry of the IPs
is such that the cyclotron emission is only present for a brief
interval, and that our short duration observations were insufficient to catch 
the cyclotron maximum. But this seems to
be the least likely explanation. If the cyclotron emission in IPs occurs
near the magnetic poles of the white dwarf, then given that the white
dwarf primaries in these systems are spinning much more rapidly than their
orbital periods (see Table 1), 
our observations were of sufficient length to have fully covered the spin
periods for most of our sources. However, if the cyclotron emission occurs
only for a brief interval during a spin period, it could be highly diluted
in cases where the spin period is very short (e.g., AE Aqr or DQ Her) relative 
to the exposure times of our observations. Examination of the individual
spectra for all of the IPs shows no significant changes from one spectrum
to the next, however, ruling this out for those sources with complete spin
phase coverage.

 Even if the cyclotron emission was 
somehow tied to the orbital period, the cyclotron emission is visible for 
$\gtrsim$ 20\% of an orbit in polars, and given that we have data for eleven 
sources, the 
chances that this emission was missed in all of them would be hard 
to accept. The eleven IPs observed by $Spitzer$ also cover a wide range in orbital
inclination, from the eclipsing system DQ Her, to the nearly face-on system
V603 Aql. It seems unlikely that beaming affects are the explanation
for our non-detection of cyclotron emission.

Discrete cyclotron harmonics are usually only seen during periods of low
mass accretion in polars (``low states''). In polar ``high states'' these 
features generally disappear and are replaced by a strong accretion
continuum. It is likely 
that the accretion rate in IPs is always equivalent to, or greater than that 
found in polar high states,
and thus emission from discrete cyclotron harmonics is suppressed. But
much of the emission from polars during high states $is$ due to cyclotron
emission, that from optically thick cyclotron emission. Examination of the
SEDs of polars in and out of high states reveals that they derive
much of their luminosity from optically thick cyclotron emission. For example, 
both AM Her (Bailey \& Axon 1981) and EF Eri (Bailey et al. 1982) in their high 
states emit strongly circularly polarized radiation. In the case of
EF Eri, discrete cyclotron harmonics are even present $during$ a high state
(Ferrario et al. 1996). Thus, while it seems likely 
that discrete harmonics might never be detected from IPs, optically thick 
cyclotron emission might occur. As detailed above, such emission would have 
been visible if it constituted more than $\sim$ 50\% of the local continuum. 

Woelk \& Beuermann (1996) note that for field strengths of B $\sim$ 5 MG, and very 
high accretion rates, {\it \.{m}} $\geq$ 1 g cm$^{\rm -2}$ s$^{\rm -1}$, 
cyclotron cooling becomes insignificant, and free-free radiation dominates. Mass
accretion rates for IPs are in this range (Buckley \& Tuohy 1989, Wu et al.
1989, Bonnet-Bidaud et al. 1982), depending on the choice for
the size of the accretion column footprint. Optically
thin free-free radiation has a spectrum of f$_{\nu}$ = $constant$, however,
and all of our spectra have power law indices that decline much more rapidly
than this. Thus, as in our arguments about optically thick cyclotron 
emission, the mid-infrared spectra show little evidence for significant
free-free emission.

The similarity of the excesses in EX Hya and V1223 Sgr to that of AE Aqr begs
the question whether synchrotron emission is the dominant cooling process
in all IPs. It should be possible to obtain adequate spectra with the IRS
for several of the systems that we have observed here using longer effective
exposure times to test this idea. It is interesting to note that both BG CMi
and DQ Her have been detected as flaring radio sources (Pavelin et al. 1994).
Perhaps a deeper survey of IPs at radio wavelengths might be warranted.

It could be that the actual field strengths for IPs lie outside the range
which would produce cyclotron emission at the mid-infrared wavelengths we
have surveyed. Sensitive searches at both visual and near-infrared wavelengths 
have failed to reveal any sign of discrete cyclotron humps. The SEDs shown
in Fig. 9 confirm this result. Thus, there are strong constraints on
optically thin cyclotron emission at optical, near-infrared, and mid-infrared 
wavelengths. In a low resolution infrared spectroscopic survey of polars 
Campbell (2007) found that rarely is there discrete cyclotron emission at 
harmonics above $n$ = 6, and only occasionally is there significant
emission at harmonics with $n$ $\leq$ 2. Assuming this is true for IPs, is
it possible for such emission to have escaped detection in the window between
2.3 and 5.5 $\mu$m? For field strengths between 7 and 9 MG, harmonics 3 through
6 would mostly occur in this unobserved bandpass, or within the water
vapor features found in the near-infrared. Given the possibility of  optically
thick emission, it seems likely that it would be easily overlooked. Thus, the 
field strengths for the program IPs could all fall 
within this narrow range. Such a proposition could be easily tested by 
obtaining phase-resolved $L$ and $M$-band photometry for a number of systems.

Alternatively, the magnetic field strengths could all be significantly higher
or lower than previously expected. 
For field 
strengths above 100 MG, the cyclotron emission would mostly
occur in the UV, and could be easily missed if dominated by the emission from
the accretion disk. But if the magnetic fields of the white dwarfs were this 
large, it would be hard to explain why these objects are not synchronously 
rotating.  We certainly cannot rule out lower field strengths. 
For field strengths of B $\leq$ 1 MG, 
very little cyclotron emission would fall within the IRS bandpass, 
and thus, could have been missed for the eleven IPs surveyed. Our
models for V1223 Sgr and EX Hya suggest the possibility of fields near 1 MG in 
those two sources. It appears that if the primary stars in IPs are magnetic, it 
is likely that their field strengths are low, B $\leq$ 1 MG.

\section{Conclusions}

We have presented a mid-infrared spectroscopic survey of eleven IPs,
and present near-infrared spectra for five of these systems. We find that, in 
general, the mid-infrared spectra of IPs are consistent
with the extension of a simple power law from the optical to the mid-infrared.
We rule out the presence of optically thin cyclotron emission to very stringent
limits, and show that optically thick cyclotron emission is an insignificant
fraction of the bolometric luminosity if it is occurring within the IRS
bandpass. We conclude that the best explanation for our results is that the
magnetic field strengths of the white dwarf primaries in intermediate polars 
are  B $\leq$ 1 MG, though we cannot rule out high accretion rate scenarios
where the cyclotron emission is a small fraction of the total luminosity. We
do report a mid-infrared excess for AE Aqr. It is likely, however, that this 
excess is due to synchrotron radiation. 

\acknowledgements This work is based in part on observations made with the 
Spitzer Space Telescope, which is operated by the Jet Propulsion Laboratory, California Institute of Technology under a contract with NASA. Support for this work was provided by NASA through an award issued by JPL/Caltech.

\begin{center}
{\bf References}
\end{center}
\begin{flushleft}
Bailey, J., \& Axon, D. J. 1981, MNRAS, 194, 187\\
Bailey, J., Hough, J. H., Axon, D. J., Gatley, I., Lee, T. J., Szkody, P., Stokes, G, \& Berriman, G. 1982, MNRAS, 199, 801\\
Bonnet-Bidaud, J. M., Mouchet, M., \& Motch, C. 1982, A\&A, 112, 355\\
Brinkworth, C. S., Hoard, D. W., Wachter, S., Howell, S., Ciardi, D. R., 
Szkody, P., Harrison, T. E., van Belle, G. T., \& Esin, A. A. 2006, AJ, 
submitted\\
Bruch, A., \& Engel, A. 1994, A\&A Supp., 104, 79\\
Buckley, D. A. H., \& Touhy, I. R. 1989, MNRAS, 344, 376\\
Campbell, R. K. 2007, PhD Thesis, in preparation\\
De Jager, O. C., Meintjes, P. J., O'Donoghue D., \& Robinson, E. L., 1994,
MNRAS, 267, 577\\
Dubus, G., Campbell, R., Kern, B., Tamm, R. E., \& Spruit, H. C. 2004, MNRAS,
349, 869\\
Ferrario, L., Bailey, J., \& Wickramasinghe, D. 1996, MNRAS, 282, 218\\
Harrison, T. E., Howell, S. B., \& Johnson, J. J. 2005b, BAAS, 37, 1276\\
Harrison, T. E., Osborne, H. L., \& Howell, S.. B. 2005a, AJ, 129, 2400\\
Harrison, T. E., Osborne, H. L., \& Howell, S.. B. 2004a, AJ, 127, 3493\\
Harrison, T. E., Howell, S. B., Szkody, P., Homeier, D., Johnson, J. J., \& 
Osborne, H. L. 2004b, ApJ, 614, 947\\
Hoard, D. W., Wachter, S., Clark, L. L., \& Bowers, T. P. 2002, ApJ, 565,
511\\
Howell, S. B., Brinkworth, C., Hoard, D. W., Wachter, S., Harrison, T. E., Chun,
H., Thomas, B., Stefaniak, L., Ciardi, D. R., Szkody, P., van Belle, G.
2006, ApJ, 646, L65\\
Meintjes, P. J., Raubenheimer, B. C., de Jager, O. C., Brink, C., Nel, H. I., North, A. R., van Urk, G., \& Visser, B. 1992, ApJ, 401, 325\\
Norton, A. J., Haswell, C. A., \& Wynn, G. A. 2004a, A\&A, 419, 1025\\
Norton, A. J., Wynn, G. A., \& Somerscales, R. V. 2004b, ApJ, 614, 349\\
Pavelin, P. E., Spencer, R. E., \& Davis, R.. J. 1994, MNRAS, 269, 779\\
Potter, S. B., Cropper, M., Mason, K. O., Hough, J. H., \& Bailey, J. A. 1997,
MNRAS, 285, 82\\
Rayner, J. T., Toomey, D. W., Onaka, P. M., Denault, A. J., Stahlberger, W. E.,
Ritter, H., \& Kolb, U. 2003, A\&A, 404, 301\\
Schenker, K., King, A. R., Kolb, U., Wynn, G. A., \& Zhang, Z. 2002, MNRAS, 337,
1105\\
Schoembs, R., \& Rehban, H. 1989, A\&A, 224, 42\\
Schwope, A. D., Beuermann, K., \&  Thomas, H. -C. 1990, A\&A, 230, 120\\
Sherrington, M. R., \& Jameson, R. F. 1983, MNRAS, 205, 265\\
Szkody, P. 1987, ApJ Supp., 63, 685\\
Vacca, W. D., Cushing, M. C., \& S. Wang 2003, PASP 115, 362\\
van Amerongen, S., Augusteijn, T., \& van Paradijs, J. 1987, MNRAS, 228, 377\\
van Amerongen, S., Kraakman, H., Damen, E., Tjemkes, S., \& van Paradijs, J.
1987, MNRAS, 215, L45\\
Vacca, W. D., Cushing, M. C., \& Rayner, J. T. 2003, PASP, 115, 389\\
Vogt, N. 1983, A\&A Supp., 53, 21\\
Warner, B. 1995, {\it Cataclysmic Variable Stars}, (Cambridge University
Press:Cambridge, England),
p367.\\
Wickramasinghe, D. T., Wu, K., \& Ferrario, L. 1991, MNRAS, 249, 460\\
Woelk, U., \& Beuermann, K. 1996, A\&A, 306, 232\\
Wu, Ci.-C., Panek, R. J., Holm, A. V., Raymond, J. C., Hartman, L. W., \& Shank,
J. H. 1989, ApJ, 339, 433\\
Wynn, G. A., King, A. R., Horne, K. 1997, MNRAS, 286, 436\\
\end{flushleft}

\begin{deluxetable}{cccccccccc}
\tablecolumns{10}
\tablewidth{0pc}
\begin{center}
\tablecaption{Observation Journal and Program Object Characteristics}
\end{center}
\tablehead{\colhead{Object} & \colhead{Obs. Date} & \colhead{U.T Start} & \colhead{SL1} & \colhead{SL2} &\colhead{LL2} & \colhead{Duration}&
\colhead{P$_{\rm orb}$} & \colhead{P$_{\rm spin}$} & \colhead{$i$\tablenotemark{1}}\\
\colhead{}&\colhead{}&\colhead{}&\colhead{[Exposure}&\colhead{time (s)}& \colhead{$\times$ \# of Cycles]}&\colhead{(min)}& \colhead{(hr)}&\colhead{(min)}&\colhead{(deg)}}
\startdata
V603 Aql & 2005-10-14 & 10:41:39 & 6 $\times$ 5 & 6 $\times$ 5 & 14 $\times$ 5 &9.5 & 3.32 & 61 & 13 \\
AE Aqr & 2004-11-13 & 08:36:52 & 14 $\times$ 2 & 14 $\times$ 2 & 30 $\times$ 3 & 12.9 & 9.88 & 0.55 & 55 \\
FO Aqr & 2005-11-23 & 11:13:52 & 240 $\times$ 2 & 240 $\times$ 3 & 120 $\times$ 5& 75.0 & 4.84 & 21 & 65 \\
TV Col & 2005-10-13 & 08:59:18 & 240 $\times$ 2 & 240 $\times$ 3 & 120 $\times$ 5 & 75.0 & 5.49 & 32 & 70 \\
TX Col & 2005-10-13 & 07:02:57 & 240 $\times$ 2 & 240 $\times$ 2 & 120 $\times$ 5& 67.0 & 5.72 & 32 & $<$ 30 \\
PQ Gem & 2006-04-23 & 20:39:48 & 240 $\times$ 2 & 240 $\times$ 2 & 120 $\times$ 4& 63.1 & 5.19 & 14 & 60 \\
DQ Her & 2005-10-19 & 23:11:49 & 240 $\times$ 2 & 240 $\times$ 3 & 120 $\times$ 5& 75.0 & 4.64 & 2.4 & 89.7 \\
EX Hya & 2006-03-04 & 01:30:01 &  60 $\times$ 2 & 60 $\times$ 2 & 120 $\times$ 5& 32.9 & 1.64 & 67 & 77 \\
GK Per & 2004-03-01 & 08:25:07 & 30 $\times$ 2  & 30 $\times$ 2 & $-$&
3.0 & 48.1 & 6.02 & $<$ 73\\
AO Psc & 2005-12-11 & 10:04:16 & 240 $\times$ 2 & 240 $\times$ 1 & 120 $\times$ 4& 50.1 & 3.59 & 13 & $<$ 30 \\
V1223 Sgr & 2006-04-17 & 11:45:43 & 60 $\times$ 5 & 60 $\times$ 5 & 120 $\times$ 5 & 50.9 & 3.37 & 61 & 13 \\
\enddata
\tablenotetext{1}{From Ritter \& Kolb (2003)}
\end{deluxetable}
\clearpage
\begin{center}
{\bf Figure Captions}
\end{center}
\begin{flushleft}
{\bf Figure 1.} The SPEX low resolution spectrum of EF Eri showing cyclotron
emission from the $n$ = 5, 6 and 7 harmonics from a 13.6 MG field in the $J$ 
and $H$ bands. The $n$ = 4 harmonic from this field falls in the $K$-band, and
is partially optically thick. Emission from He I (+ H I Pa $\gamma$) at 1.083 
$\mu$m is clearly present. \\

{\bf Figure 2.} The IRS spectra of the intermediate polars. a) the spectra
of TV Col, TX Col and PQ Gem. The spectra for TV Col and TX Col have been 
vertically offset by 0.005 and 0.003 mJy, respectively, for clarity. 
The spectra have also been color-coded to delineate the different IRS modes: 
blue is SL2, green is SL1, and red is LL2. b) the data for EX Hya, V603 Aql and 
V1223 Sgr. The spectra for EX Hya and V603 Aql have been offset by 0.016 and 
0.0035 mJy, respectively. The spectrum of V603 Aql appears to have a strong 
emission line at 8 $\mu$m. c) the data for FO Aqr, AO Psc and DQ Her. The 
spectra for FO Aqr and AO Psc have been offset by 0.0068 and 0.003 mJy, 
respectively. d) The IRS data for GK Per.

{\bf Figure 3.} The IRS spectrum of AE Aqr. While the short wavelength data
are consistent with a power law, at $\lambda$ $>$ 10 $\mu$m, there is a
considerable excess. The solid line in this figure is a power law (f$_{\nu}$
$\propto$ $\nu^{\rm 0.9}$) + blackbody (T$_{\rm eff}$ = 140 K) model. Emission
lines from the Humprheys $\alpha$ and Pfund $\alpha$ (+ Hu $\beta$) lines of H I
are clearly present.\\

{\bf Figure 4.} The IRTF SPEX spectrum of V603 Aql. The spectrum is dominated
by emission from the Paschen and Brackett series of H I, and strong He I emission 
lines (labeled) on a blue continuum. V603 Aql is believed to
be a low inclination system, consistent with the line profiles, though there
is a blue asymmetry on some of the stronger lines. There is no evidence
for the presence of a secondary star in these data.\\

{\bf Figure 5.} The SPEX data for FO Aqr. The He I lines (e.g., that at 1.0830
$\mu$m) are weaker than those in the spectrum of V603 Aql, and the line 
profiles are doubled, consistent with an orbital inclination of $i$ = 
65$^{\circ}$. \\

{\bf Figure 6.} The IRTF spectrum of DQ Her. DQ Her is a faint source, 
$K$ $\sim$ 13.1, and a challenging target at this spectral resolution. Thus, 
the data have substantially lower S/N, and the spectrum shown here was boxcar
smoothed by a factor of 5. The H I and He I lines are broader than
the previous two objects, consistent with DQ Her's edge-on inclination
($i$ = 89.7$^{\circ}$).\\

{\bf Figure 7.} The SPEX data for AE Aqr are dominated by the secondary 
star--even the Na I doublet at 0.82 $\mu$m is clearly visible. The atomic 
absorption features are consistent with a spectral type of K4, except that 
such an object should have strong CO absorption features. There is no evidence
for $^{\rm 12}$CO absorption (positions for the main bandheads indicated) in 
this spectrum, suggesting that carbon is highly deficient (the absorption
feature at 2.38 $\mu$m is actually a doublet of Mg I).  \\

{\bf Figure 8.} The IRTF spectrum of GK Per is also dominated by its late-type
subgiant secondary star. As discussed by Harrison et al. (2005b), nearly
all of the atomic absorption features in the spectrum of GK Per are weaker
than they should be, suggesting a sub-solar metallicity for the secondary
star.\\

{\bf Figure 9.} a) The optical to mid-infrared SEDs for the program
objects ($JHK$ photometry from Hoard et al. 2002, unless otherwise noted): FO 
Aqr (filled 
circles, $UBVJHK$ from Szkody 1987), PQ Gem (open circles, $UBVRI$ from 
Potter et al. 1997), AO Psc (stars, $UBV$ from Bruch \& Engel 1994), TV Col 
(triangles, $UBV$ from Bruch \& Engel 1994) and DQ Her (squares, $UBVRI$ from
Schoembs \& Rehban 1988).  Each of the SEDs has been offset for clarity.  
Power law spectra (dashed lines) have been fitted to the SEDs of FO Aqr 
(f$_{\nu}$ $\propto$ $\nu^{1.0}$) and AO Psc (f$_{\nu}$ $\propto$ $\nu^{1.1}$).
 b) the SEDs for TX Col (solid triangles, $UBVRIJHK$ from Buckley \& Touhy 1989),
V603 Aql (filled circles, $UBV$ from Bruch \& Engel 1994)), 
EX Hya (open circles, $UBV$ from Vogt 1983), V1223 Sgr (stars, $UBVRI$ from
Bonnet-Bidaud et al. 1982), and GK Per (triangles, $UBVJHK$ from Sherrington
\& Jameson 1983).  Power laws have been fit to the SEDs of V603 Aql (f$_{\nu}$ 
$\propto$ $\nu^{1.0}$) and V1223 Sgr (f$_{\nu}$ $\propto$ $\nu^{1.3}$). The 
SED of EX Hya shows a near-infrared excess which is probably due to its 
late-type secondary star. The SED of GK Per is dominated by its secondary 
star. Both V1223 Sgr and EX Hya appear to have a long wavelength excess. 
Error bars are plotted when larger than the symbol size. \\

{\bf Figure 10} a) Model cyclotron spectra for a 3 MG field with a plasma
temperature of T$_{\rm pl}$ = 20 keV, a viewing angle of $\Theta$ = 57${\circ}$, and for a large
range of (from top to bottom) log($\Lambda$) = 5.0, 3.0, 1.0, $-$3.0, and
$-$5.0. As the optical depth is increased, the discrete cyclotron harmonics
($n$ = 1 is the right-most) are replaced with a nearly featureless continuum.
b) cyclotron + power law models fit to the IRS data for EX Hya. Error bars 
are plotted for the middle SED. The cylotron models shown here have B = 3 MG,
T$_{\rm pl}$ = 20 keV, $\Theta$ = 57${\circ}$, and log($\Lambda$) = 1.0, 3.0, 
and 5.0.  The cyclotron models were normalized so that the peak of the $n$ = 
4 harmonic, located at 9 $\mu$m, accounts for 20\% of the power law flux at 
this wavelength. c) the same, but for a field strength of 5 MG. The
cyclotron spectra were normalized so that the peak of the $n$ = 3 harmonic
at 7.6 $\mu$m accounts for 20\%  of the power law flux at this wavelength.\\

{\bf Figure 11.} The SED of AE Aqr. The optical/near-infrared fluxes of AE Aqr
are dominated by its secondary star. $UBVRIJHKL'M'$, 11.7 and 17.6 $\mu$m 
fluxes are from Dubus et al. (2004), while the 25 and 60 $\mu$ fluxes are from
IRAS. The IRS spectrum is the thin line, and a power law (f$_{\nu}$ $\propto$ $\nu^{0.9}$) + cyclotron model
(B = 1 MG, T$_{\rm pl}$ = 20 keV, $\Theta$ = 57$^{\circ}$, and log($\Lambda$) = 
3.0) is shown as the heavy green line.\\
\end{flushleft}
\begin{figure}
\plotone{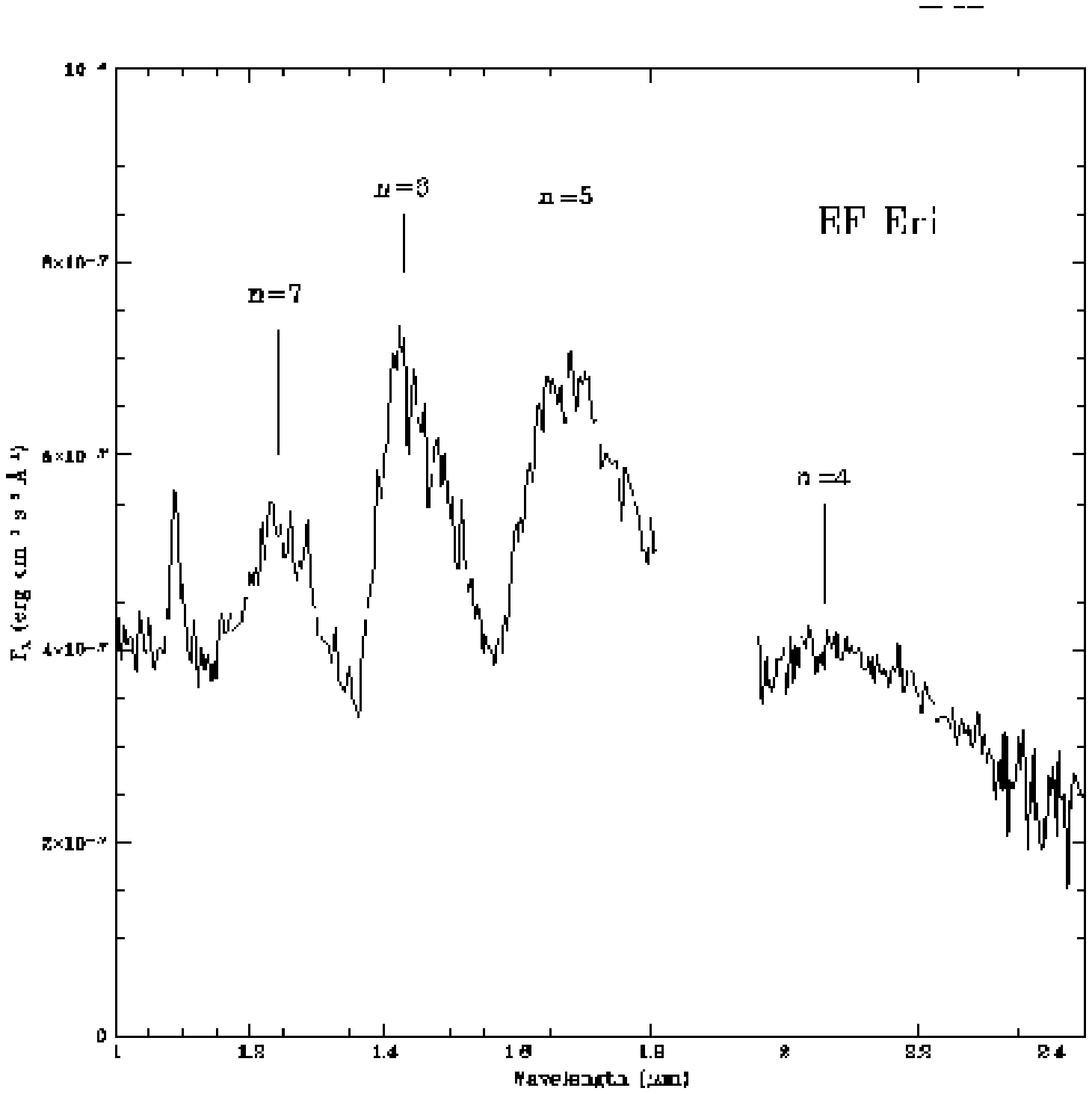}
\end{figure}
\begin{figure}
\plotone{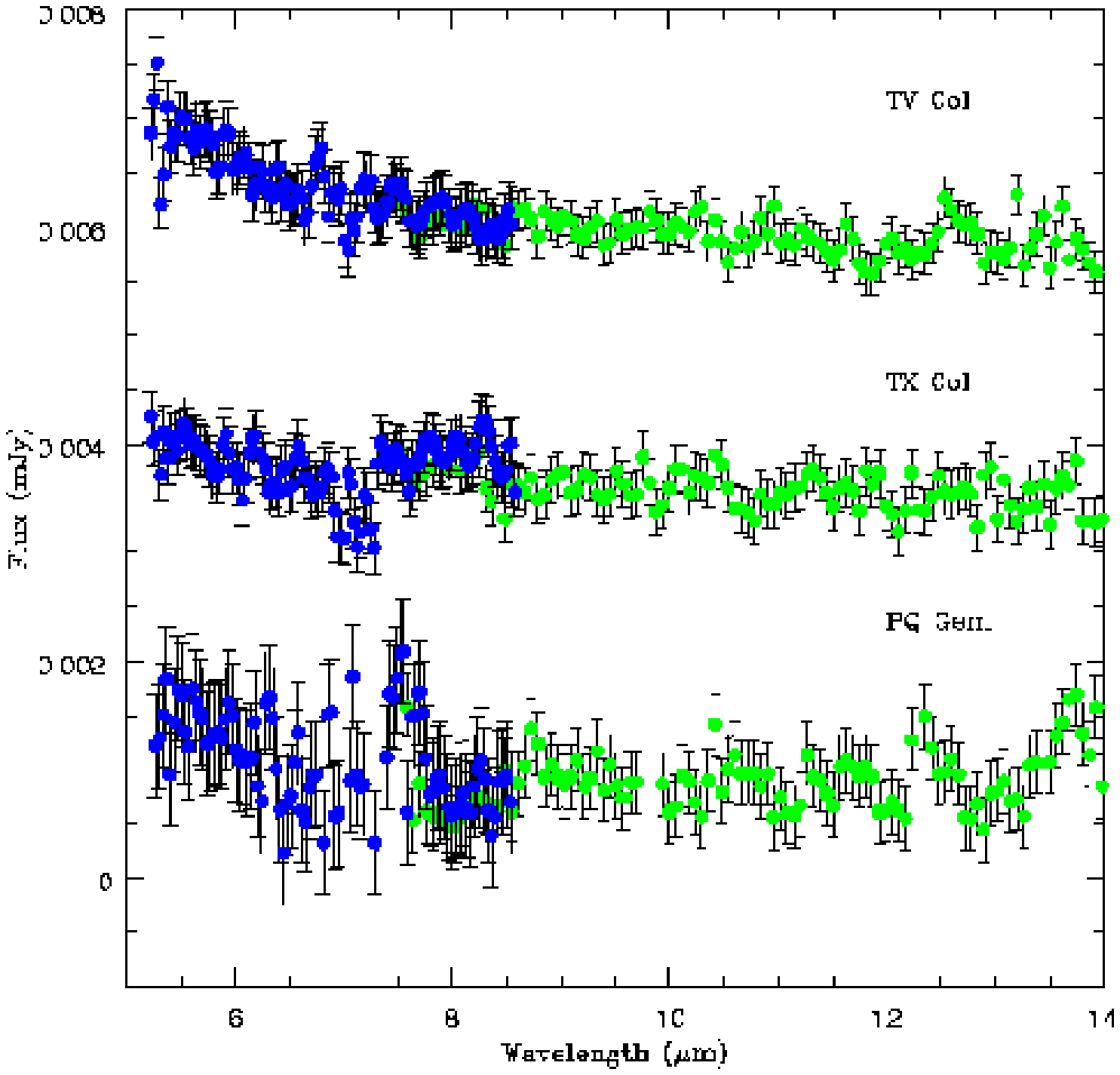}
\end{figure}
\begin{figure}
\plotone{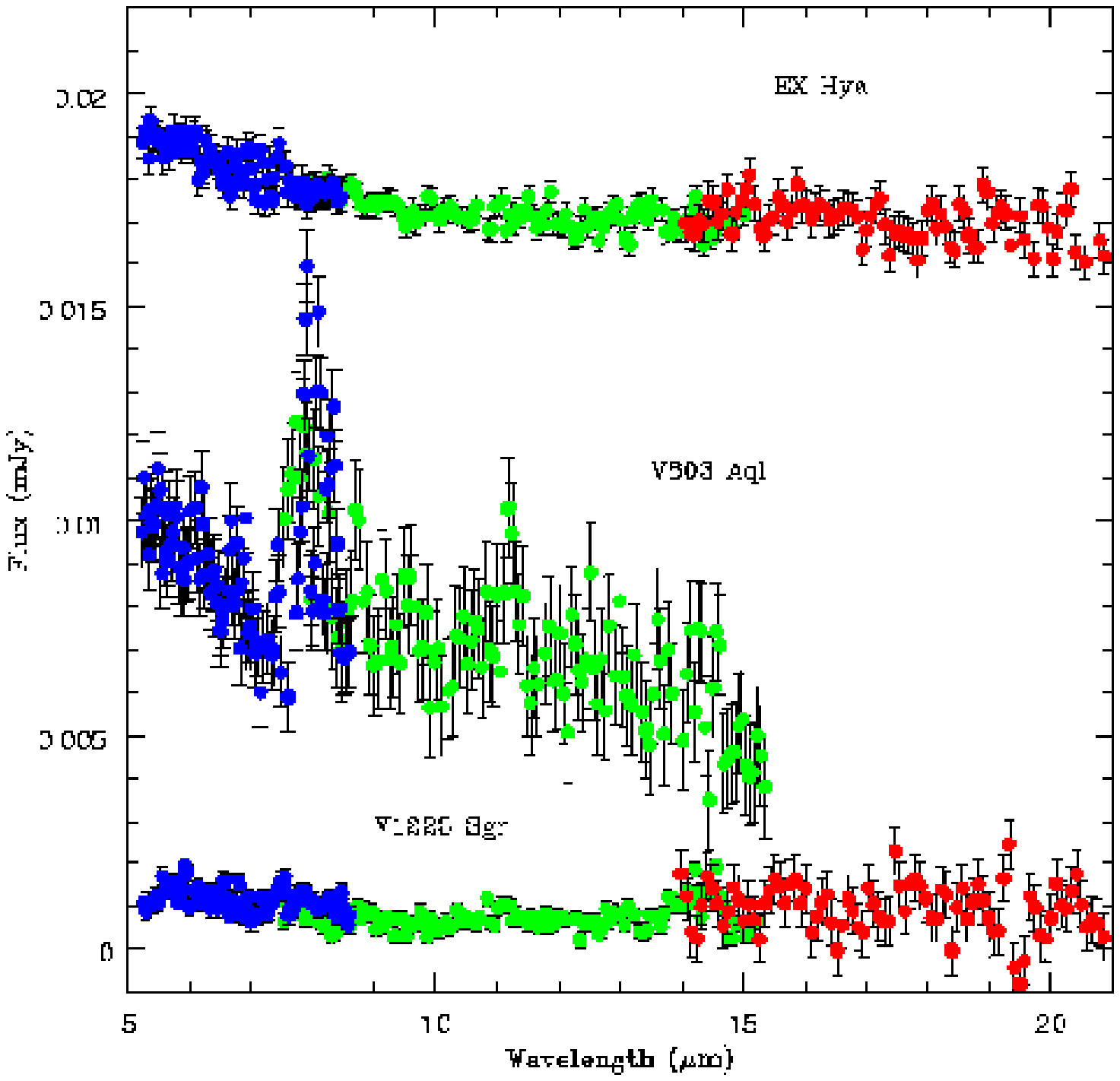}
\end{figure}
\begin{figure}
\plotone{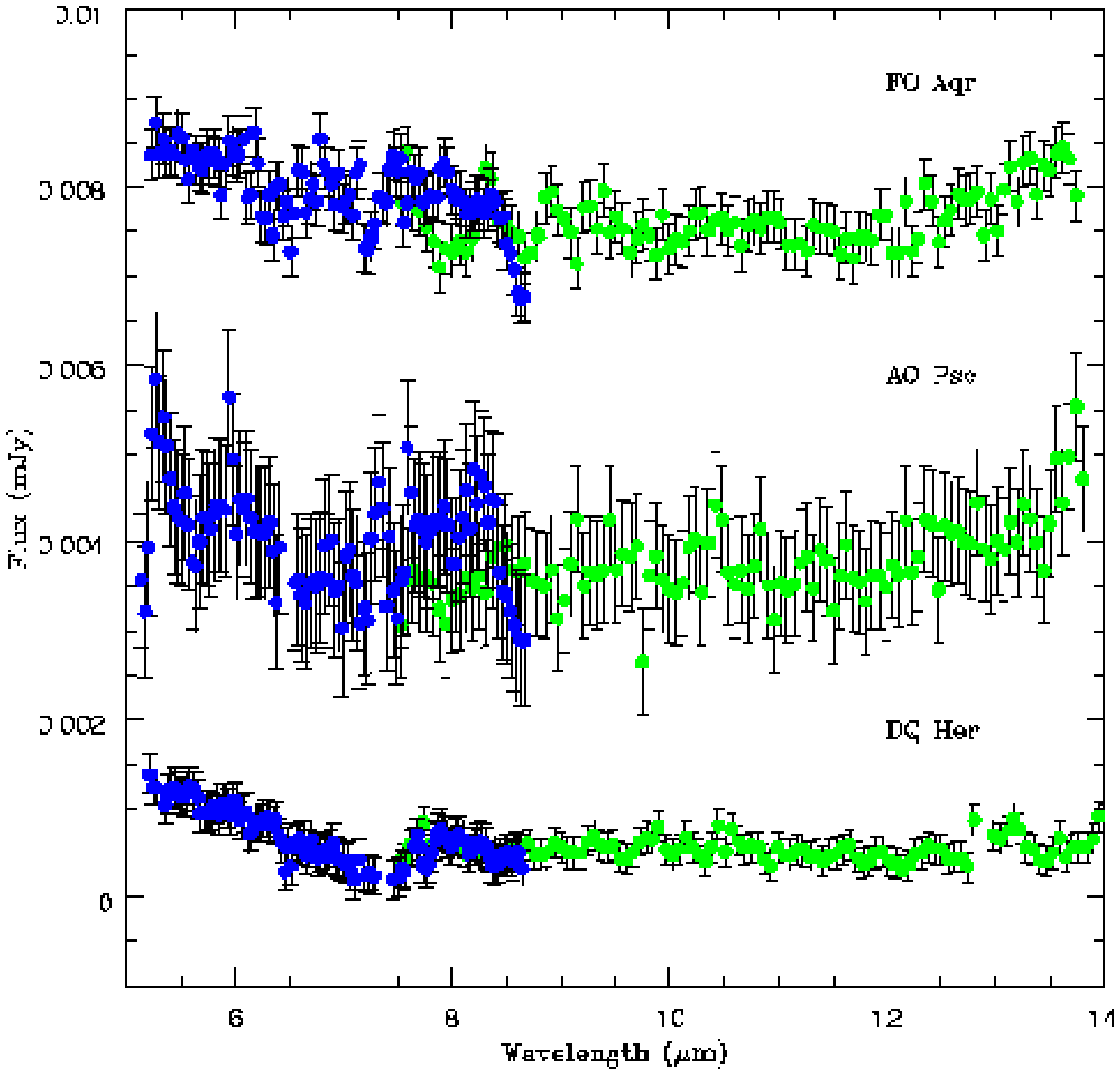}
\end{figure}
\begin{figure}
\plotone{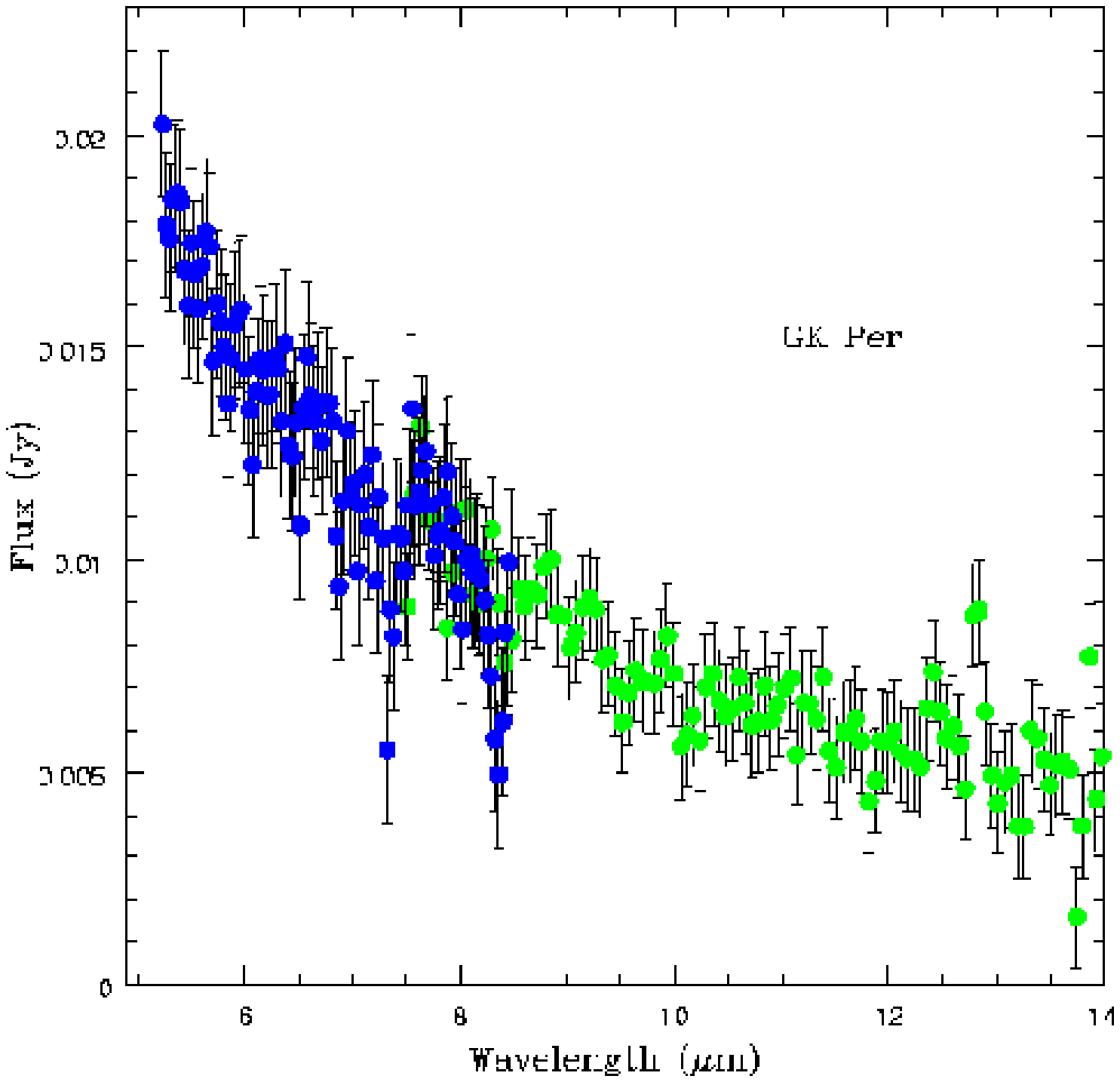}
\end{figure}
\begin{figure}
\plotone{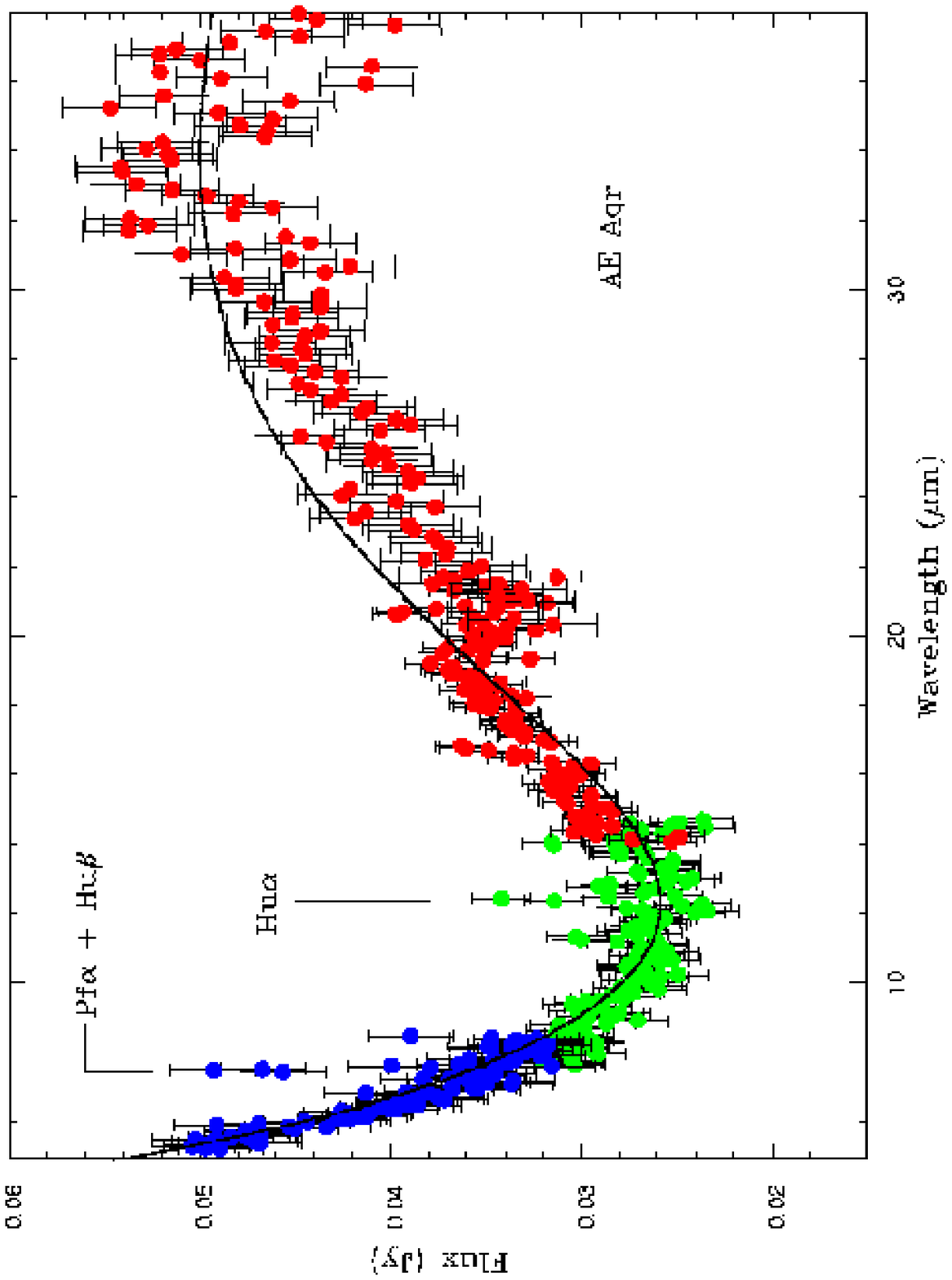}
\end{figure}
\begin{figure}
\plotone{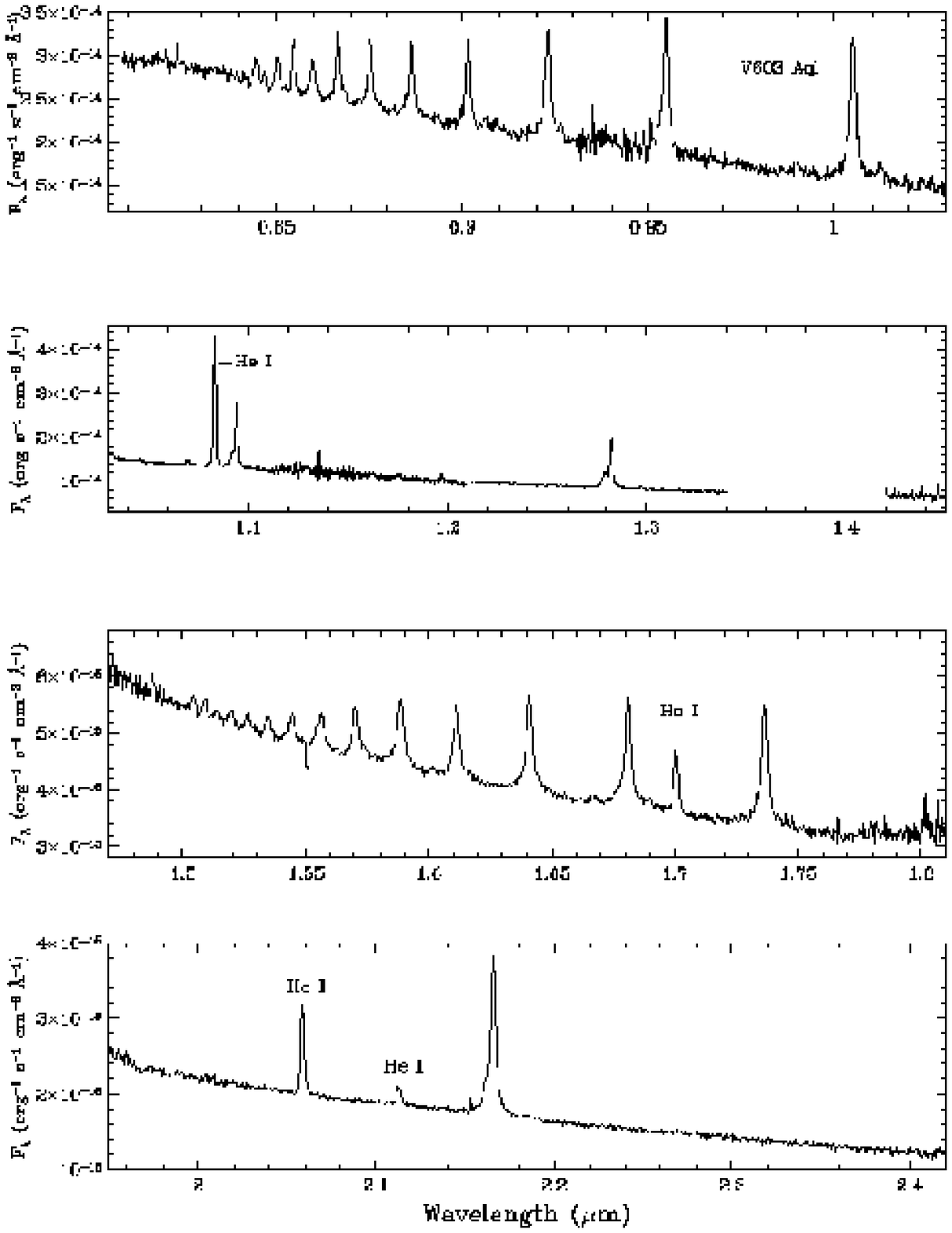}
\end{figure}
\begin{figure}
\plotone{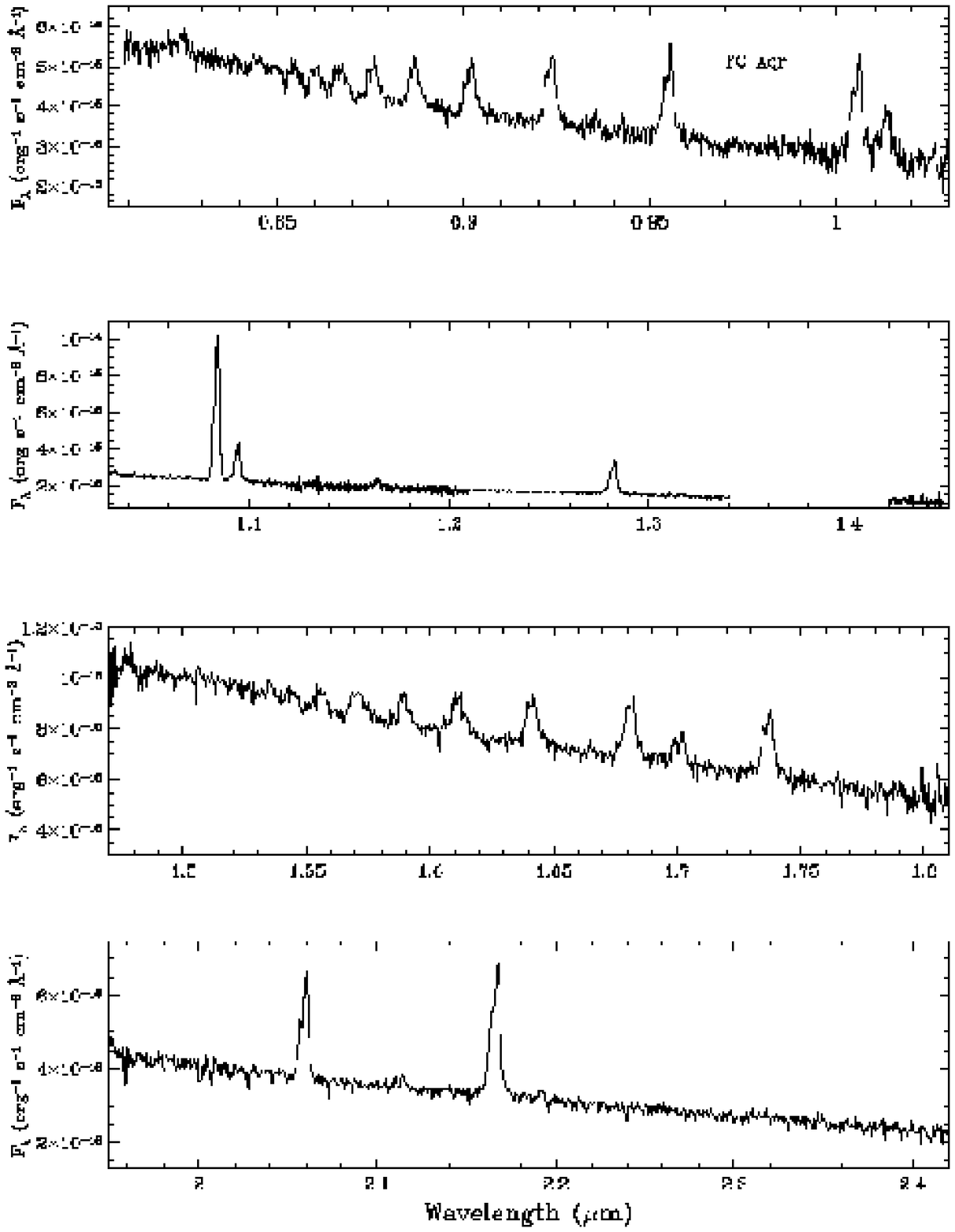}
\end{figure}
\begin{figure}
\plotone{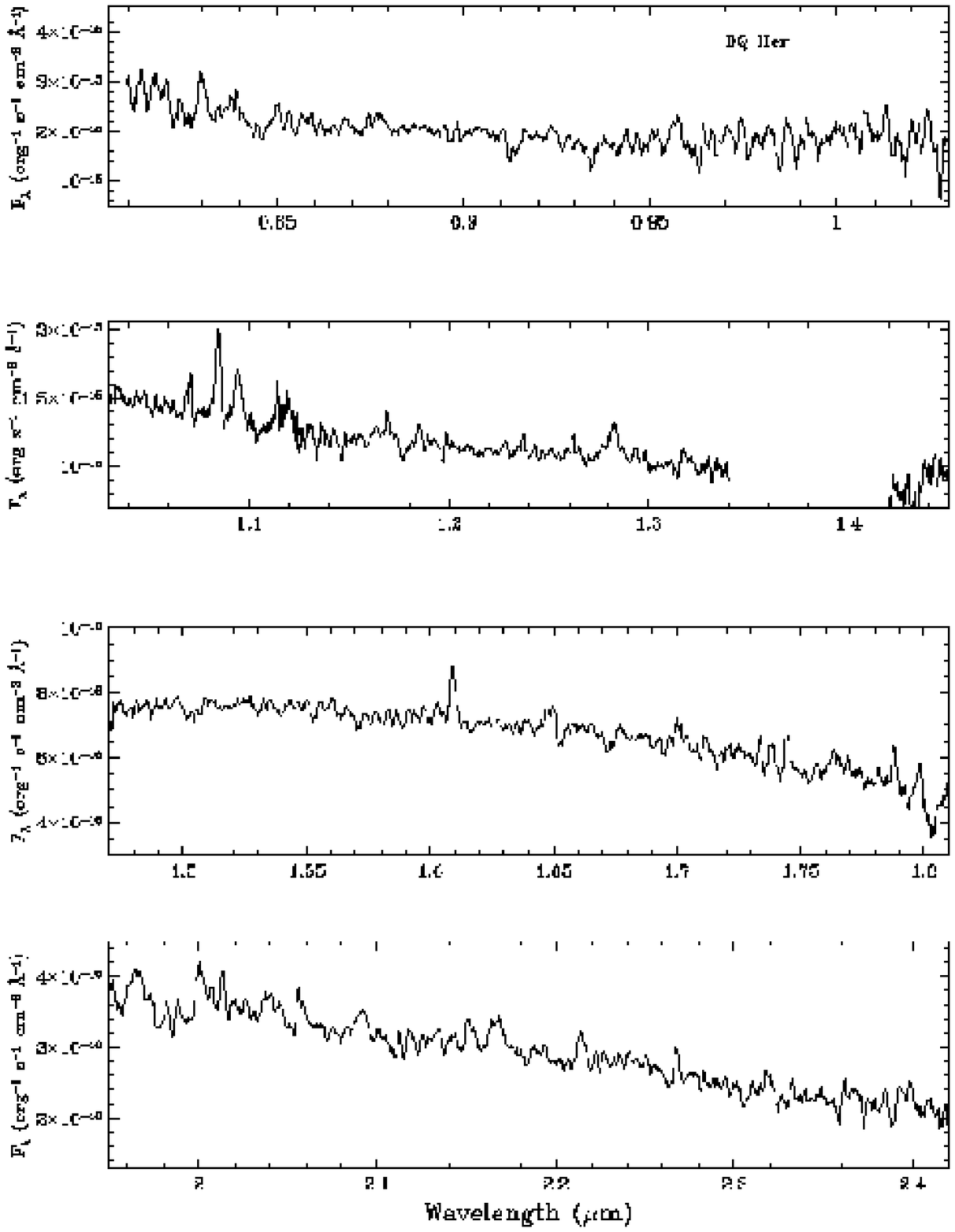}
\end{figure}
\begin{figure}
\plotone{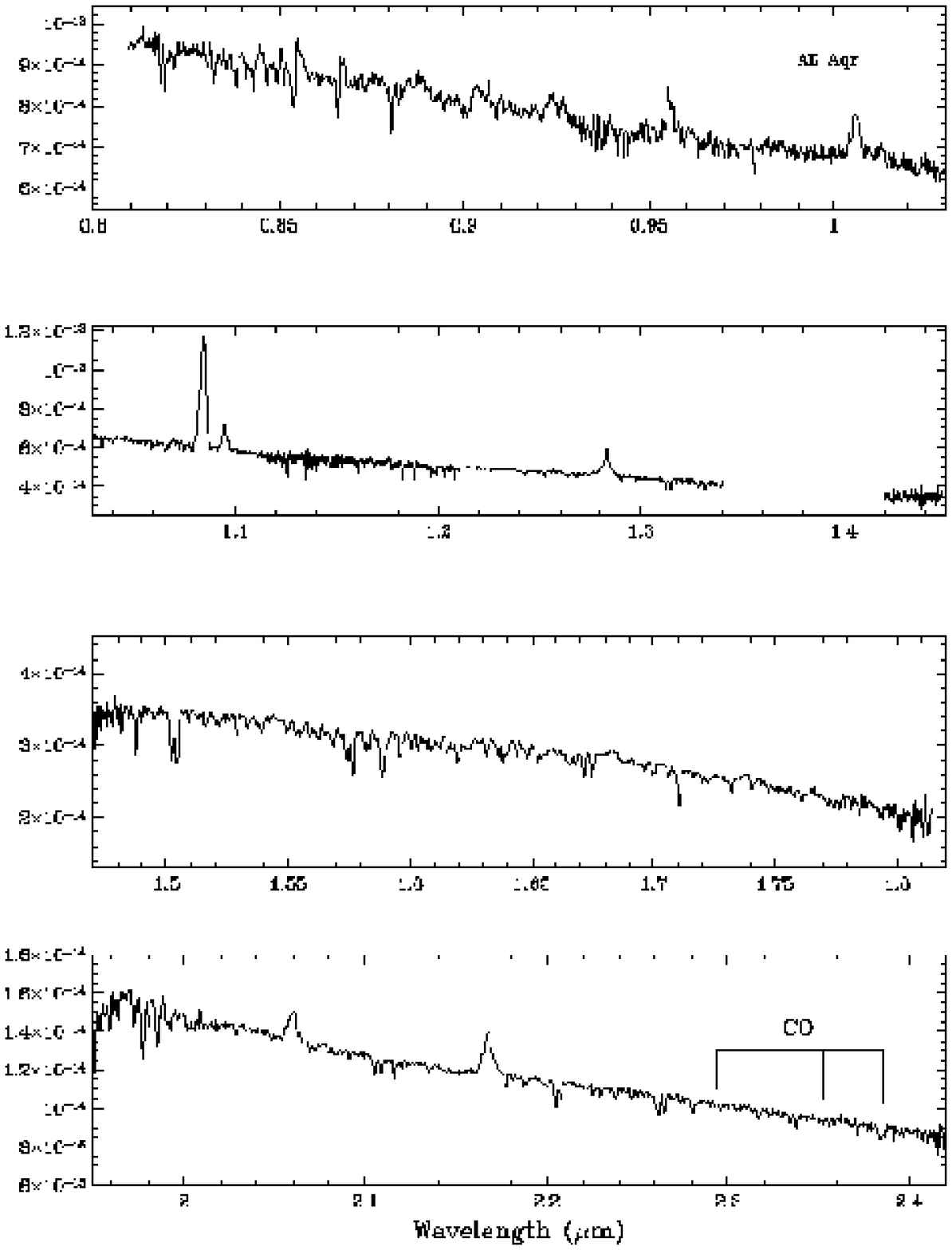}
\end{figure}
\begin{figure}
\plotone{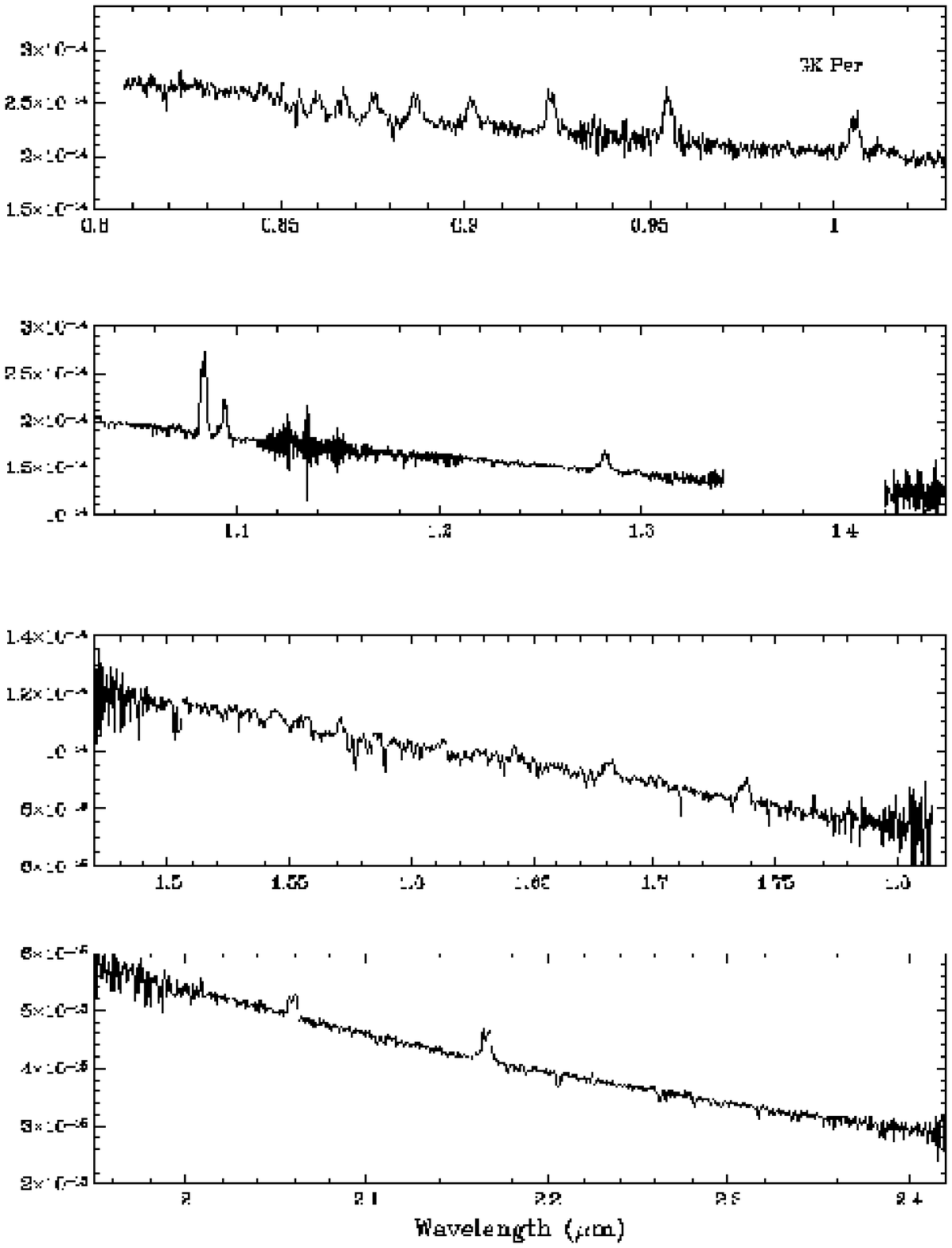}
\end{figure}
\begin{figure}
\plotone{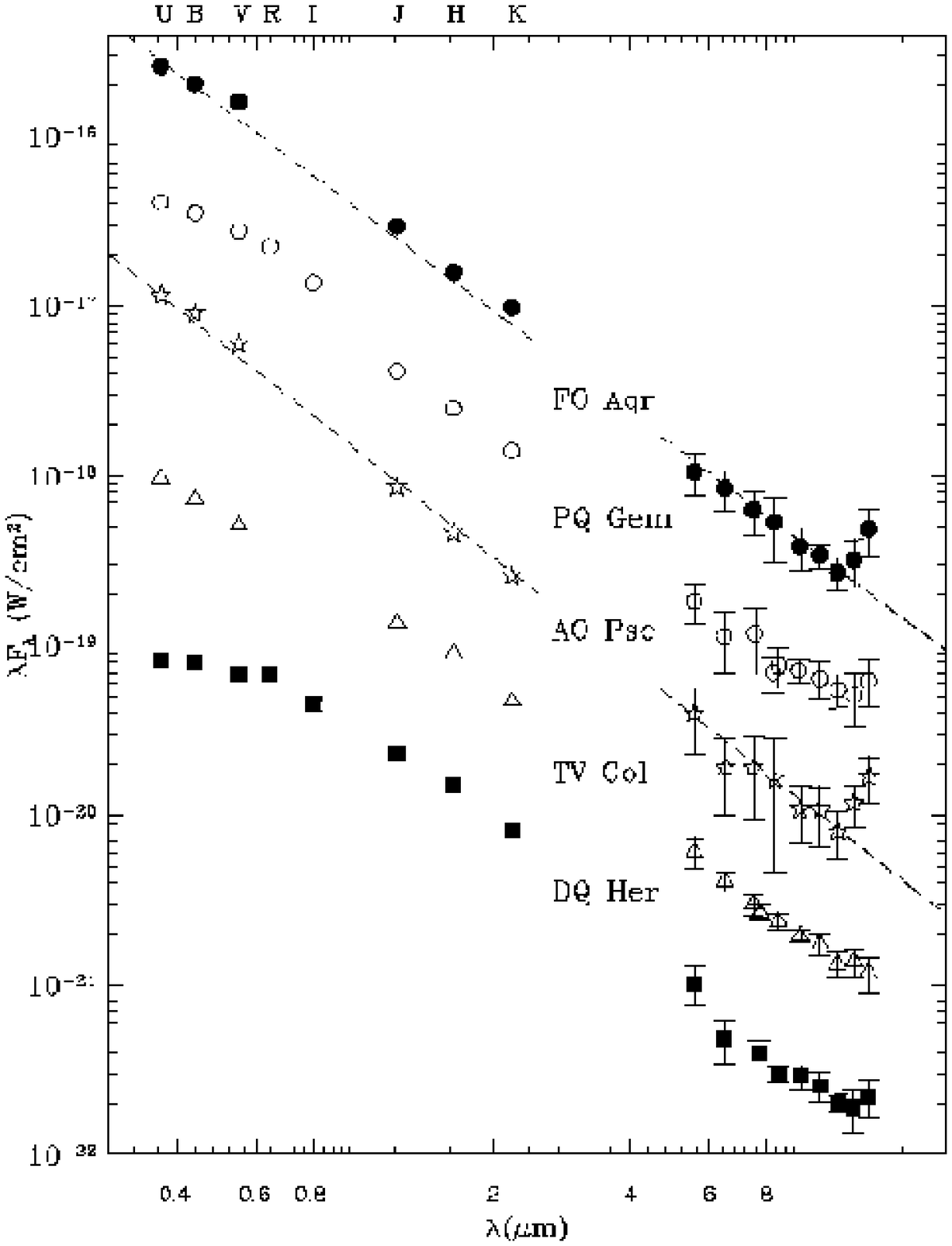}
\end{figure}
\begin{figure}
\plotone{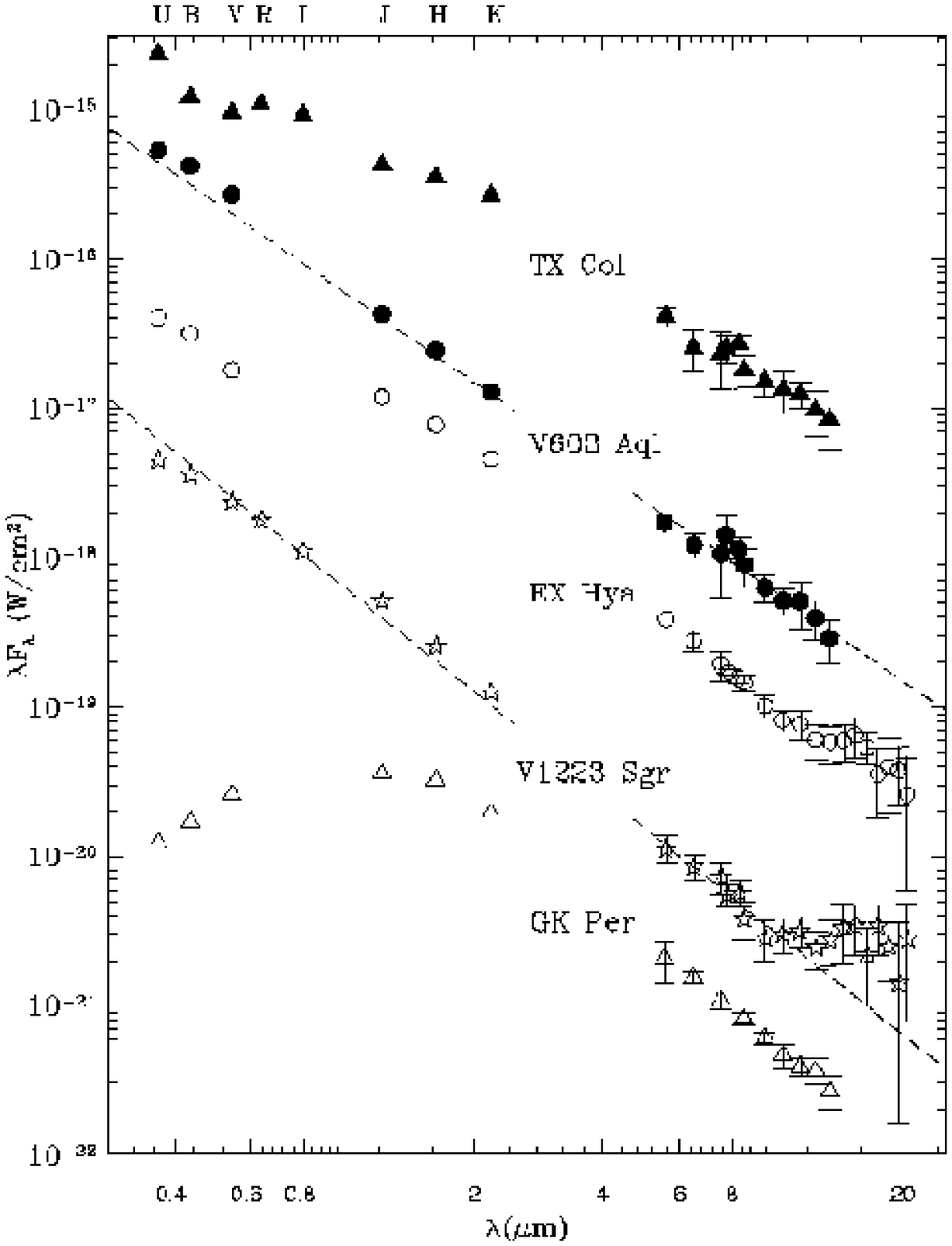}
\end{figure}
\begin{figure}
\plotone{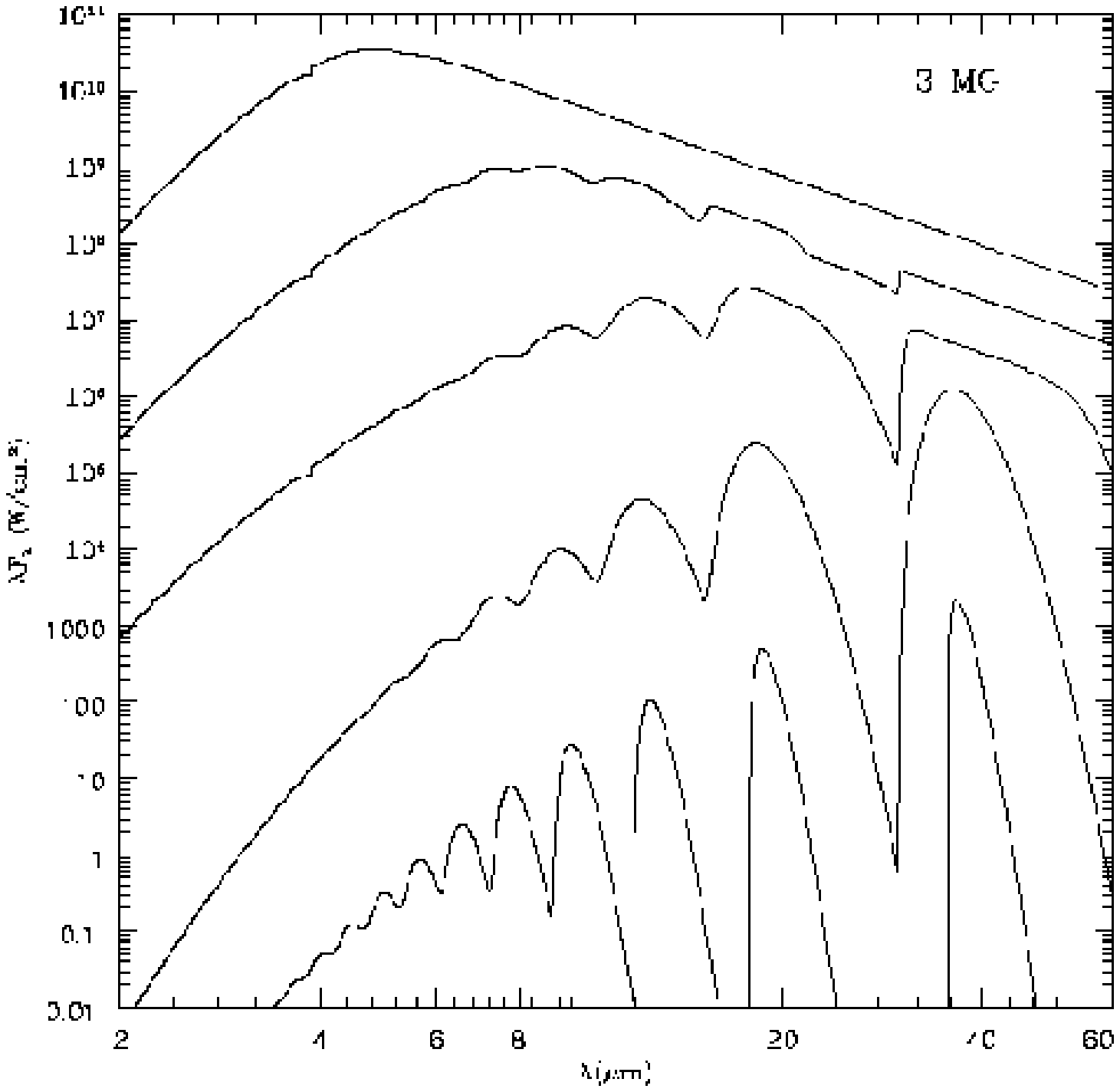}
\end{figure}
\begin{figure}
\plotone{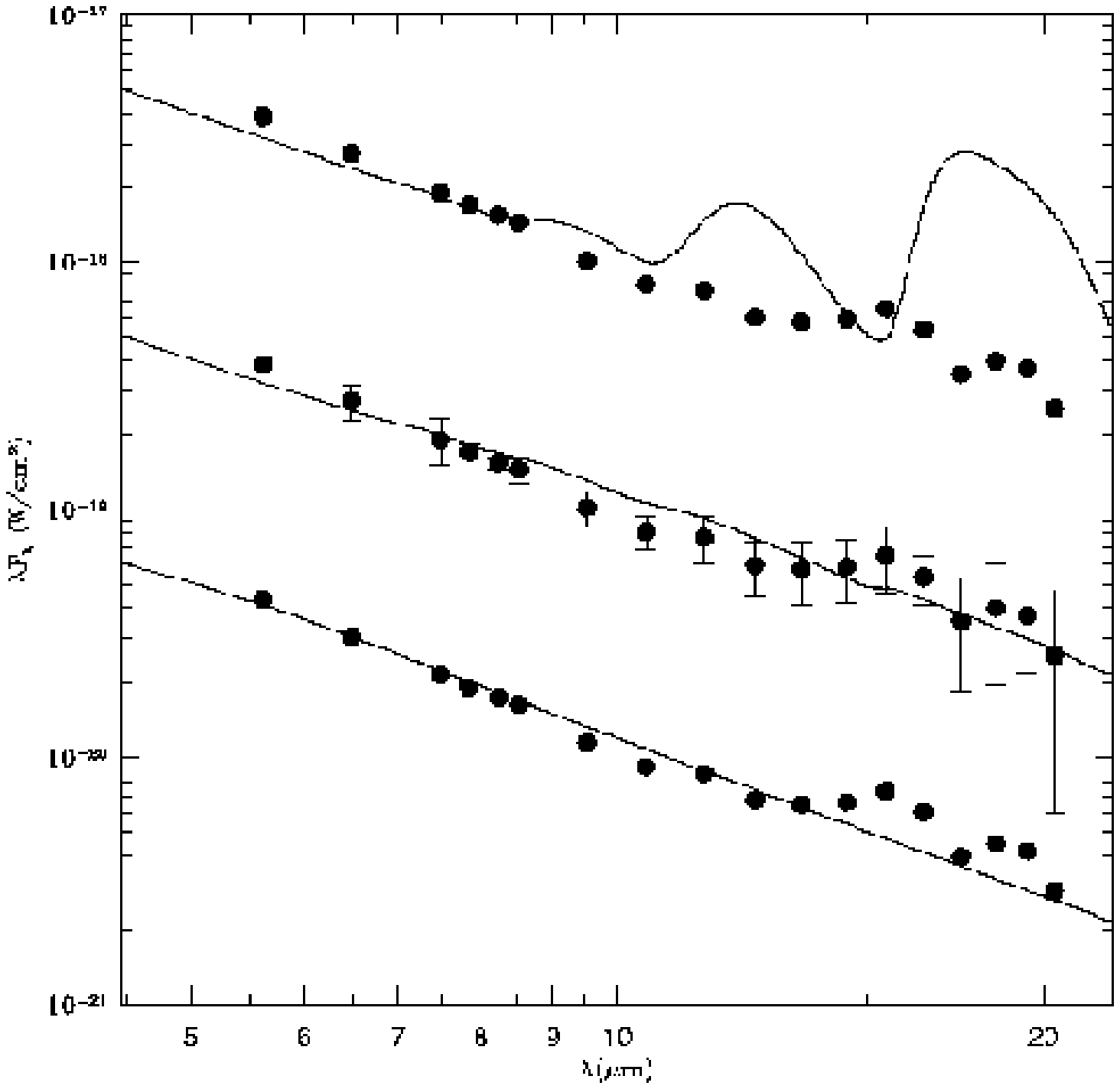}
\end{figure}
\begin{figure}
\plotone{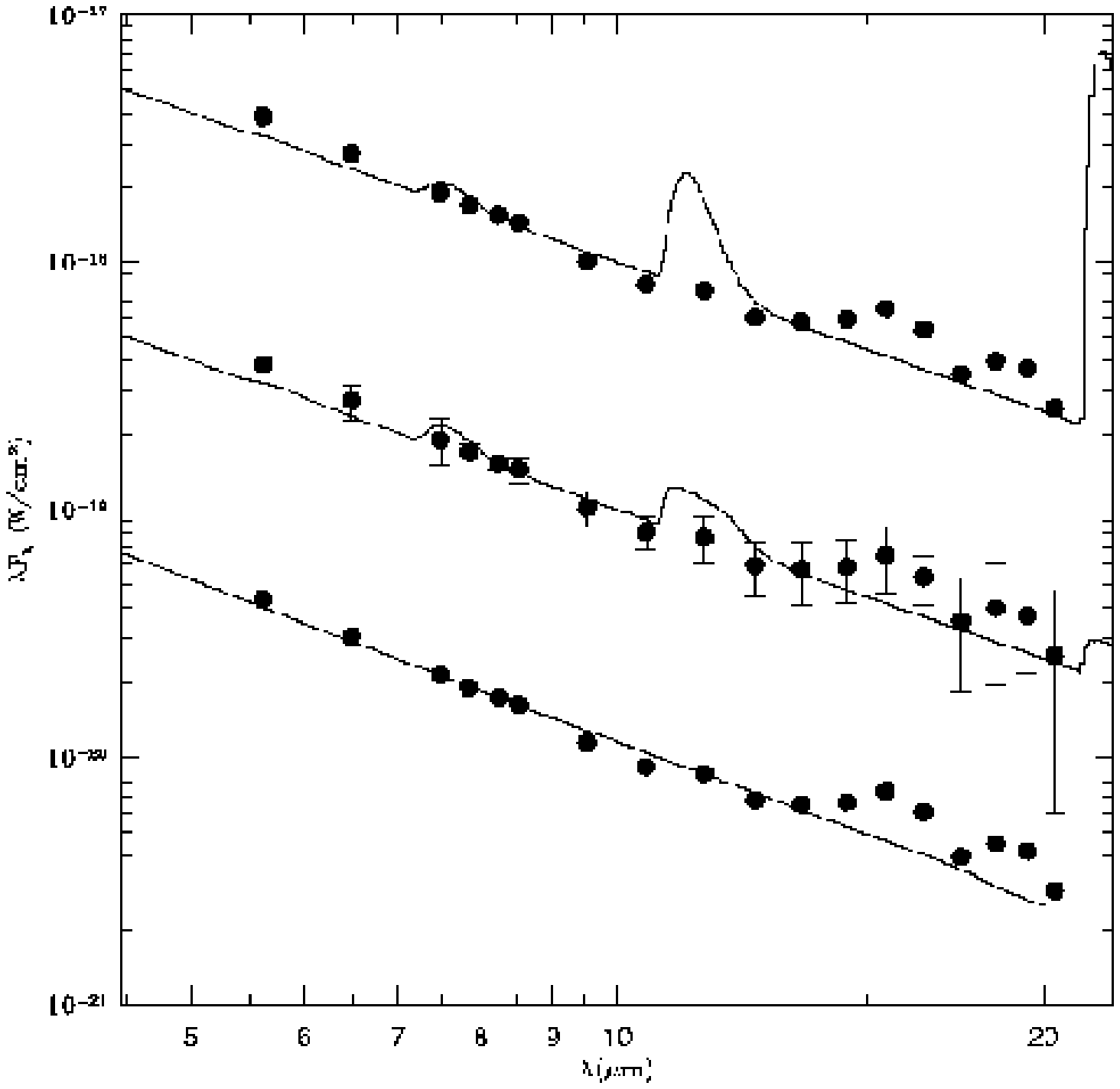}
\end{figure}
\begin{figure}
\plotone{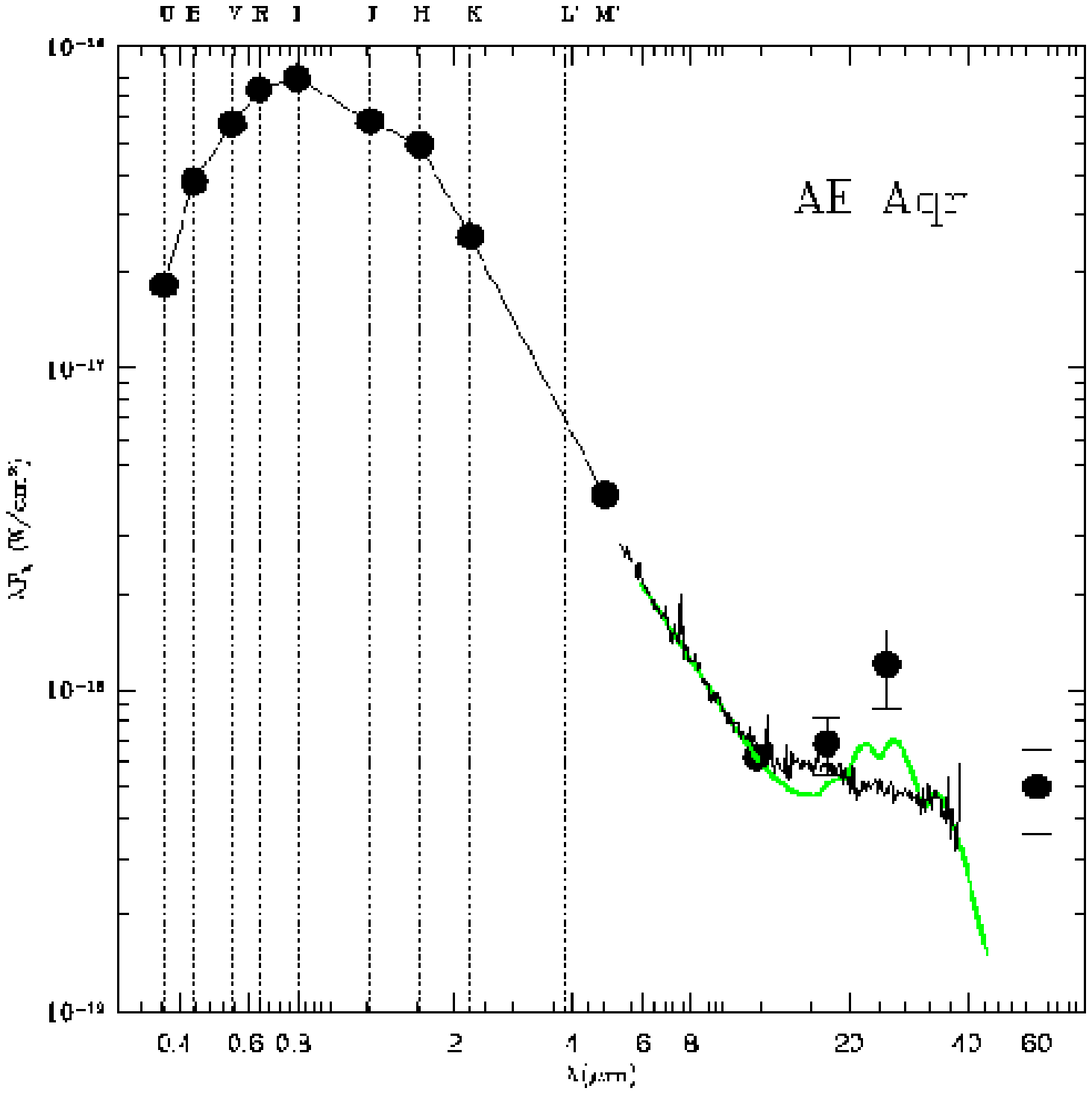}
\end{figure}

\end{document}